\documentclass[a4paper,12pt,oneside]{article}
\usepackage[latin1]{inputenc}
\usepackage{amsmath}
\usepackage{amsfonts}
\usepackage{amssymb}
\usepackage{graphicx}
\usepackage{footnote}
\usepackage{float}
\usepackage{subfigure}
\DeclareGraphicsExtensions{.bmp,.png,.pdf,.jpg,.eps}

\usepackage[colorlinks]{hyperref}
\hypersetup{
    colorlinks=true,
    linkcolor=blue,
    filecolor=magenta,  
    urlcolor=[rgb]{0.00,0.00,0.50},    
    citecolor=red,
    }

\usepackage{url}

\advance \textheight by \topskip
\usepackage{amsmath}
\usepackage[english]{babel}
\makeatletter
 \@addtoreset{equation}{section}
\makeatother

\usepackage{amsmath}
\numberwithin{equation}{section}

\advance \textheight by \topskip
\usepackage{amsmath}
\usepackage[english]{babel}

\def\be{\begin{equation}}
\def\ee{\end{equation}}

\def\A{\mathbb A}

\def\Z{\mathbb Z}
\def\R{\mathbb R}

\def\P{\mathbb P}

\def\N{\mathbb N}

\usepackage{todonotes}

\begin{document}
\title{
{\bf Hidden symmetries of rationally deformed superconformal mechanics \\
 }
}
\author{{\bf Luis Inzunza and Mikhail S. Plyushchay} 
 \\
[8pt]
{\small \textit{ Departamento de F\'{\i}sica,
Universidad de Santiago de Chile, Casilla 307, Santiago 2,
Chile  }}\\
[4pt]
 \sl{\small{E-mails: 
 \textcolor{blue}{luis.inzunza@usach.cl},
\textcolor{blue}{mikhail.plyushchay@usach.cl}
}}
}
\date{}
\maketitle

\begin{abstract}
We study the spectrum generating closed nonlinear superconformal
algebra that  describes $\mathcal{N}=2$ super-extensions 
of rationally deformed  quantum harmonic oscillator 
and  conformal mechanics models with coupling constant 
$g=m(m+1)$, $m\in {\mathbb N}$.
It has a nature of a nonlinear  finite $W$ superalgebra being
generated by higher derivative integrals, 
and  generally  contains  several different copies of either deformed superconformal  
$\mathfrak{osp}(2|2)$ algebra
in the case of superextended rationally deformed conformal mechanics  models, or 
deformed super-Schr\"odinger algebra in the case of super-extension
of rationally deformed harmonic oscillator systems.
\end{abstract}

\section{Introduction}

The development of factorization method  \cite{fac} 
closely related to the method of Darboux transformations \cite{Darboux,MatSal} and 
supersymmetric mechanics \cite{Witten,Cooper} 
allows  essentially extend the list of
exactly  solvable quantum mechanical systems. 
A particular class of the systems which can be solved  following this approach
are given by multi-soliton reflectionless potentials \cite{reflectionless}, 
the Calogero model  \cite{Calogero1,Calogero2,Perel}
 and its $\mathcal{P}\mathcal{T}$-regularizations \cite{PT1}.
All these quantum mechanical systems are intimately 
related with completely integrable classical field theories. Another 
class of such  systems corresponds to rational deformations (extensions) 
of the quantum harmonic oscillator (QHO) and conformal mechanics model 
of de Alfaro, Fubini and Furlan (AFF) 
 \cite{AFF,WeiJor,Gomez,Grandati2,CarPly,CarPly2,CIP}. 
This extremely large variety of new exactly solvable systems 
is very interesting in the light of hidden symmetries study
\cite{Cariglia}. 
Each of them possesses  some 
peculiar
characteristics,
which can appropriately be  described in terms of
higher order, sometimes explicitly depending on time,
integrals of motion.

These new families of solvable systems can be generated from 
the free particle, the harmonic oscillator and the AFF model
which, together with their supersymmetric extensions,
are characterized by  (super)conformal symmetry  \cite{SCM1,SCM2,IvaKriLev2,SCM3,RJAA,Olaf}. 
This symmetry plays  important role in the description of various physical 
 aspects and phenomena
such as non-relativistic holography and AdS/CFT correspondence
\cite{DTson,KBJM,MT,CDPH},
geodesic motion in background of charged black holes \cite{CDKKTV,AIPT,GibTow}, 
quark confinement \cite{BTGE},  and  Bose-Einstein condensates 
\cite{BEC1,BEC2},
to name a few.

For quantum systems with multi-soliton potentials 
and  those  corresponding to 
rational extensions of  the harmonic oscillator and AFF models, 
the whole picture is complicated by the presence of hidden symmetries.   
For example,  the spectrum of  multi-soliton potential systems  
is divided into finite number of  bound states and a 
continuous part.
These spectral peculiarities together with reflectionless nature of the systems are reflected by the
higher order Lax-Novikov integral of motion.
The presence of this integral 
leads to  
extension of the $\mathcal{N}=2$ super-Poincar\'e symmetry 
of such systems to the exotic nonlinear $\mathcal{N}=4$
supersymmetry, in which the Lax-Novikov integrals of 
the super-partners compose  the central charge
of superalgebra \cite{CJNP,AraMatPly,AraPly}.  
Another example where the  Lax-Novikov integral plays important 
role is provided by  the Calogero model  \cite{CorLecPly}. 
In the case of its $\mathcal{PT}$-regularized two-particle version
being perfectly invisible zero-gap quantum system, 
this higher order integral generates  a nonlinear  extension 
of the (super)conformal algebra \cite{JM1,JM2}.

On the other hand, any exactly solvable system belonging to the 
family of the QHO and AFF systems with rationally extended potentials  has an
infinite discrete spectrum with a finite (possibly zero) number of missing energy levels
in its lower part.  The infinite equidistant  tower of discrete states in such systems 
can be considered as analog of the conduction band in the spectrum 
of finite-gap quantum systems, while grouped and separated by gaps  discrete levels 
in the lower part of the spectrum  can be treated as analogs of the valence bands.
Instead of Lax-Novikov integral, any such  a system  is characterized  by the presence 
of  at least two pairs of higher order ladder operators which form   
 the complete spectrum generating sets:  they allow to obtain an arbitrary physical state from  
another \cite{CarPly2,CIP}. These higher order ladder operators also 
encode the information on the number of valence bands (or, number of gaps in the spectrum), 
as well as  the total number of  states in them. It also was observed 
that each separate pair of these ladder operators generate 
 a non-linear deformation of the conformal algebra \cite{CIP}. 
 Having this observation in mind, the problem on which we work in this paper is to find the complete 
spectrum generating algebra for this family of the systems.

Nonlinear conformal algebras 
appeared earlier  in the context of  $W$ symmetries \cite{JYN,W}.
Symmetries described by finite $W$ algebras 
appear, particularly, in anisotropic harmonic oscillator systems with commensurable 
frequencies \cite{BDKL},  and  in the Coulomb problem in  spaces of
constant curvature, that corresponds to the so called  Higgs oscillator and its
generalizations \cite{Higgs,Zhe,Evnin}. 

In this work, we are dealing with rationally extended Hamiltonians 
which can be obtained via factorization method 
based on the  iterative Darboux-Crum-Krein-Adler (DCKA)
mapping  \cite{Darboux,MatSal,Crum,Krein,Adler}
applied to  the QHO  eigenstates.  
In particular, 
we are interested in those cases in which two different 
selections of seed states produce the same, but
modulo  a nonzero  discrete shift, system.
One choice of the seed states  for DCKA transformation 
consists in selecting  only physical eigenstates,
while  the second choice corresponds  to selection  
of a certain ``complementary"  set of non-physical states 
of harmonic oscillator with negative energy
produced by the  
spatial Wick rotation 
 $x\rightarrow ix$ of physical states. 
This is what we name the  Darboux duality property \cite{CIP}. 
As is well known, any Darboux (or DCKA) transformation allows to construct
the $\mathcal{N}=2$ superextended system described
by linear (or nonlinear) Poincar\'e superalgebra.
The presence of another such a  transformation with the described properties
allows us to extend superalgebra generally 
to a nonlinearly extended superconformal algebra,
which, via the repeated (anti)-commutation relations,
creates  the sets of the ladder operators of the partner systems 
which compose new bosonic generators, and new 
fermionic generators of the superalgebra.  In this way  
we finally  obtain the closed spectrum generating 
nonlinear superalgebra of the  $\mathcal{N}=2$ superextended system.
The resulting closed nonlinear superalgebra contains generally 
several different copies of either  deformed superconformal  $\mathfrak{osp}(2|2)$ algebra
in the case of superextended rationally deformed AFF models, or 
deformed super-Schr\"odinger algebra in the case of super-extension
of rationally deformed harmonic oscillator systems.

The rest of the paper is organized as follows. 
In Section \ref{Dar}, we  recall how  in general case 
nonlinear $\mathcal{N}=2$ super-Poincar\'e symmetry
is generated by DCKA transformations.  
Using this construction, in Section \ref{isoharconf}
we show how the requirement of inclusion of 
the spectrum generating sets of operators expands 
 linear Poincar\'e supersymmetry
of  the $\mathcal{N}=2$  superextended  AFF model and harmonic oscillator
up to superconformal $\mathfrak{osp}(2|2)$ and 
super-Schr\"odinger symmetries, respectively.
Interesting peculiarity is that for the AFF model 
the $\mathcal{N}=2$ supersymmetry can be realized in both  
unbroken and broken phases,  for which the complete sets of 
the true and time-depending  integrals are related by an automorphism 
of the superconformal $\mathfrak{osp}(2|2)$ 
symmetry. In the case of the harmonic oscillator, 
linear $\mathcal{N}=2$ super-Poincar\'e symmetry 
is realized only in the unbroken phase.
The mechanism of generation of superconformal and super-Schr\"odinger 
symmetries in  the $\mathcal{N}=2$  superextended  AFF model and harmonic oscillator
is generalized in Section \ref{gen} for  rational deformations 
of these systems, where, as we shall see, 
nonlinearly extended 
versions of the indicated symmetries appear.
Concrete simple examples illustrating  the general 
construction are presented
 in Section \ref{examples}. Last Section \ref{conclusion}
 is devoted to the discussion and outlook. 
 Several Appendices provide necessary technical 
 details for  the main text.

\section{Darboux transformations and supersymmetry}
\label{Dar}

The procedure we will apply  is based on the  Darboux-Crum-Adler-Krein transformations \cite{Darboux,MatSal,Crum,Krein,Adler}.
Here, we summarize  its basic ingredients.

Consider a Hamiltonian operator
$L= 
-\frac{d^2}{dx^2}+V(x)$, and choose
the set of its eigenstates $\psi_1,\ldots,\psi_n$ of eigenvalues 
$\lambda_1,\ldots,\lambda_n$
 so that 
their Wronskian  $W$ takes nonzero values on the domain of 
the quantum system $L$.
In such a way we generate a new, super-partner  quantum system 
\be
\label{DarbouxHamil}
\breve{L}=- 
\frac{d^2}{dx^2}+\breve{V}(x)\,,\qquad
\breve{V}(x)=V(x)-
2
\big(\ln(W(\psi_1,\psi_2,\ldots,\psi_n)\big)''\,,
\end{equation}
defined on the same domain as the initial system $L$.
An arbitrary solution $\psi_\lambda$ to the second order differential equation
$L\psi_\lambda=\lambda\psi_\lambda$,
$\lambda\neq \lambda_j$, $j=1,\ldots,n$,  
is  mapped into the wave function 
\begin{equation}
\label{Darbouxstates}
\psi_{[n],\lambda}=\frac{W(\psi_1,\psi_2,\ldots,\psi_n,\psi_\lambda)}{W(\psi_1,\psi_2,\ldots,\psi_n)}\,,
\end{equation}
which is an
 eigenstate of $\breve{L}$ of the same eigenvalue,
$\breve{L}\psi_{[n],\lambda}=\lambda\psi_{[n],\lambda}$, 
and has  the same  nature of  physical or non-physical
state  as the pre-image $\psi_\lambda$. 
{}From the pair $L$ and $\breve{L}$, 
we construct the $\mathcal{N}=2$ superextended  system 
described by the $2\times 2$ matrix Hamiltonian  and the supercharges
given by
\be\label{Hlambda*}
\mathcal{H}=
\left(
\begin{array}{cc}
H_1\equiv  \breve{L}-\lambda_*&    0 \\
0 & H_2\equiv  L-\lambda_*     
\end{array}
\right),\qquad
\mathcal{Q}_1=
\left(
\begin{array}{cc}
  0&    \A_n \\
\A_n^\dagger &  0     
\end{array}
\right),
\qquad 
\mathcal{Q}_2=i\sigma_3\mathcal{Q}_1\,.
\ee
The system also is characterized by the integral $\Sigma=\frac{1}{2}\sigma_3$. 
Here  $\lambda_*$ is a constant, 
while  $\A_n$ and $\A_n^\dagger$
are the operators defined recursively by
relations
\begin{equation}\label{Andef}
\A_n=A_nA_{n-1}\ldots A_1,\qquad 
A_{k}= 
\A_{k-1}\psi_{k}\frac{d}{dx}\frac{1}{\A_{k-1}\psi_{k}},\qquad k=1,\ldots,n,
\end{equation}
where $A_0=1$ is assumed. 
These operators intertwine the partner Hamiltonians,
$\A_n L=\breve{L}\A_n$,
$\A_n^\dagger \breve{L}=L\A_n^\dagger$.  They
provide an alternative representation 
for relation (\ref{Darbouxstates}),
$\psi_{[n,\lambda]}=\A_n\psi_\lambda$,
and generate the polynomials in  Hamiltonian operators,
\begin{eqnarray}\label{An+An-}
&\A_n^\dagger \A_n=\prod_{j=1}^n(L-\lambda_j)\equiv P_n(L)\,,\qquad
\A_n \A_n^\dagger  =P_n(\breve{L})\,.&
\end{eqnarray}
Up to (inessential for us here) multiplicative factor, we also have a relation
$\A_n^\dagger \psi_{[n,\lambda]}=\psi_\lambda.$
By the construction, the kernel of $\A_{n}$ 
is spanned  by $n$ seed states $\psi_j$, $j=1,\ldots,n$,
while 
$\ker \A_{n}^\dagger=\text{span}\,\{\A_{n}\widetilde{\psi_1},\ldots,\A_{n}\widetilde{\psi_n}\}$, where
\begin{equation}
\label{secondlinearindependent}
\widetilde{\psi_\lambda(x)}=\psi_\lambda(x) \int^x \frac{d\xi}{(\psi_\lambda(\xi))^2}\
\end{equation} 
is a linearly independent solution to the 
equation
$L{\psi}=\lambda{\psi}$
corresponding to the same value of $\lambda$,
$L\widetilde{\psi_\lambda}=\lambda\widetilde{\psi_\lambda}$.

Operator $\Gamma=\sigma_3$,
$\Gamma^2=1$,  plays a role 
of the  $\Z_2$-grading operator that
identifies $\mathcal{H}$ and  $\Sigma$ 
as even (bosonic) generators of the 
superalgebra,
$[\Gamma,\mathcal{H}]=0$, $[\Gamma,\Sigma]=0$,
while  $\mathcal{Q}_a$ are identified 
as  odd (fermionic) 
generators,   
$\{\Gamma,\mathcal{Q}_a\}=0$. 
The integrals satisfy the (anti)-commutation relations 
\be\label{N2susy}
[\mathcal{H},\mathcal{Q}_a]=0\,, \quad
[\mathcal{H},\Sigma]=0\,,\quad
\{\mathcal{Q}_a,\mathcal{Q}_b\}=2\delta_{ab}P_n(\mathcal{H}+\lambda_*)\,,
\quad
[\Sigma,\mathcal{Q}_a]=-i\epsilon_{ab}\mathcal{Q}_b
\ee
of the $\mathcal{N}=2$ superalgebra
which has a linear Lie-superalgebraic nature in the case of $n=1$ and is nonlinear
for $n>1$. Integral  $\Sigma$ is a generator of a $U(1)$ 
$\mathcal{R}$-symmetry.

We will apply this construction to the 
AFF model as well as to the QHO.
The peculiarity of these systems which we will exploit 
is that 
under Wick rotation $x\rightarrow ix$ 
their Hamiltonian operators  transform in a simple way 
just by multiplying by minus one. 
 As a consequence, under such a transformation
 solution to the second order
 differential equation $L\psi(x)=\lambda \psi(x)$ 
 transforms into the function $\psi(ix)$ which satisfies 
 the equation  $L\psi(ix)=-\lambda \psi(ix)$.
This property allows us to
produce from the AFF as well as from the QHO models 
the pairs  of exactly  the same super-partner systems
but with the  mutually shifted spectra  
by choosing  alternative  sets  of the seed states. 
As a result,
the $\mathcal{N}=2$ supersymmetric structure 
of the corresponding extended systems related to 
the AFF model will
expand upto superconformal $\mathfrak{osp}(2| 2)$
symmetry in the case of $n=1$, while
for $n>1$ the corresponding 
systems will be described by some nonlinear
extension of the $\mathfrak{osp}(2| 2)$.
When $L$ corresponds to
the QHO, 
the $\mathfrak{osp}(2|2)$ will expand up to the 
super-Schr\"odinger symmetry and 
its nonlinear extensions in the cases of $n=1$ and $n>1$, respectively.
Essential  ingredients in superconformal  extensions 
will be  the  sets of the spectrum generating ladder operators 
 \cite{CarPly2,CIP}.

\section{Superconformal and super-Schr\"odinger symmetries}
\label{isoharconf}

In this section we review and compare the  superconformal symmetries 
of the $\mathcal{N}=2$ superextended AFF and quantum harmonic 
oscillator models. The 
approach exposed here 
will form a basis  for the analysis of nonlinearly extended 
superconformal symmetries 
of  rational deformations of these systems.

\subsection{Intertwining   and ladder operators}

The dynamics of a particle 
in a  harmonic oscillator potential 
is described by the Hamiltonian operator   
$L= 
-\frac{d^2}{dx^2}+x^2  
\equiv L_0$ 
with
spectrum 
$E_n= 
2n+1
$,
$n=0,1,\ldots,$
and (nonnormalized) eigenstates  
$\psi_{n}(x)=H_n(x)e^{-x^2/2}$, where $H_n(x)$ 
are the Hermite polynomials. 
The choice of the ground state $\psi_0(x)=e^{-x^2/2}$ as the seed state for 
the Darboux transformation generates according to (\ref{Andef})  the first order 
differential operators
$A_1=\frac{d}{dx}+x\equiv a^-$, and $A_1^\dagger\equiv a^+$. 
In this case the partner system  (\ref{DarbouxHamil}) 
is just the shifted QHO
$\breve{L}=L_0+2$ with the shift constant equal to the 
distance between neighbour energy levels
in its spectrum.
 Because of this, 
the intertwining operators $a^-$ and $a^+$ are nothing else
that the harmonic oscillator's  ladder operators 
satisfying relations $a^+a^-=L_0-1$, $a^-a^+=L+1$,
and so, $[a^-,a^+]=2$, and 
$[L_0,a^\pm]=\pm 2a^\pm$.
If instead of taking the ground state 
as the seed state for generation of the Darboux transformation
we choose a non-physical eigenstate 
$\psi_0(ix)=e^{x^2/2}$ of eigenvalue $-1=-E_0$,
we obtain the same ladder operators 
as the intertwining-factorizing operators $A_1=-a^+$  and $A^\dagger_1=-a^-$.
In this case, however,  their relation to operators 
$A_1$ and $A_1^\dagger$ is interchanged, and consequently 
$\breve{L}=L_0-2$.  This difference will 
be essential to expand  the $\mathcal{N}=2$ supersymmetry 
of the superextended QHO up to 
its dynamical super-Schr\"odinger symmetry. 
\vskip0.1cm

The AFF model with a confining harmonic potential term, sometimes also 
called an isotonic oscillator \cite{WeiJor}, 
is described by the Hamiltonian operator
\begin{equation}
\label{isoham}
L^{iso}_\nu=
-\frac{d^2}{dx^2}+x^2+\frac{\nu(\nu+1)}{x^2}
\,,\qquad \nu\in\R\,,
\end{equation}
defined on the domain 
$\{ \psi\in L^2((0,\infty),dx)\vert \psi(0^+)=0\}$.
The case $\nu=0$ is understood  as a limit  $\nu\rightarrow 0$
which corresponds to the half-harmonic oscillator with the infinite potential well  put at $x=0$.
For $\nu\geq-1/2$ 
the spectrum of the system and corresponding (non-normalized) 
eigenstates are 
\be\label{Enun}
E_{\nu,n}=
2\nu+4n+3, \qquad 
\psi_{\nu,n}=x^{\nu+1}\mathcal{L}_n^{(\nu+1/2)}(x^2)e^{-x^2/2}\,,\qquad
n=0,1,\ldots,
\ee
where  $\mathcal{L}_{n}^\alpha$ 
are the generalized Laguerre polynomials  \cite{Perel}. 
The coupling constant $g(\nu)=\nu(\nu+1)$ 
has a symmetry $g(\nu)=g(-\nu-1)$.
Applying the  transformation $\nu\rightarrow -\nu-1$
to  eigenstates (\ref{Enun}), one obtains 
 $\psi_{-\nu-1,n}=x^{-\nu}\mathcal{L}_n^{(-\nu-1/2)}(x^2)e^{-x^2/2}$,
  which satisfy relation $L^{iso}_\nu\psi_{-\nu-1,n}=(-2\nu+4n+1)\psi_{-\nu-1,n}$.
  For $\nu<0$ these are  eigenfunctions (\ref{Enun}) 
of the system $L^{iso}_{\vert \nu\vert -1}$.
On the other hand,  for $\nu>0$ these are non-physical eigenstates
of $L^{iso}_{ \nu}$,
  which are non-normalizable singular at $x=0^+$  functions.
In spite of the non-physical nature,  such states will play important role
in our constructions.
 In the limit  $\nu \rightarrow 0$, 
system  (\ref{isoham}) transforms into the 
half-harmonic oscillator with the infinite potential barrier at $x=0$, 
which we will denote below by $L_{0^+}$. 
In accordance with the relations $H_{2n+1}(x)=2(-4)^nn!x\mathcal{L}_n^{(1/2)}(x^2)$ and 
$H_{2n}(x)=(-4)^nn!\mathcal{L}_{n}^{(-1/2)}(x^2)$, 
physical eigenstates $\psi_{\nu,n}$ of the system (\ref{isoham}) 
in the limit $\nu\rightarrow 0$ 
take the form of the odd states of the quantum harmonic 
oscillator $L_0$ being eigenstates of $L_{0^+}$ on its domain,
while the indicated non-physical eigenstates 
transform in this limit into  the even eigenstates of $L_0$, which do not    
satisfy the Dirichlet boundary condition for the 
half-harmonic oscillator system $L_{0^+}$.
 
 Let us take as the initial system $L=L_{\nu-1}^{iso}$ 
 with $\nu\geq1/2$ 
 and choose its ground state $\psi_{\nu-1,0}=x^\nu e^{-x^2/2}$
 to generate 
 the Darboux transformation. In this way 
 we obtain the first order differential operators
 \be \label{A1nu-def}
 A_1=\frac{d}{dx}+x-\frac{\nu}{x}\equiv A^-_\nu\,,\qquad
 A_1^\dagger =-\frac{d}{dx}+x-\frac{\nu}{x}\equiv A^+_\nu\,,
 \ee
 for which $\text{ker}\,A^-_\nu=\psi_{\nu-1,0}$ and 
 $\text{ker}\,A^+_\nu=(\psi_{\nu-1,0})^{-1}$.
 They satisfy relations 
 \begin{eqnarray}\label{AA+nu}
 &A^+_\nu A^-_\nu=L_{\nu-1}^{iso}-2\nu-1\,,\qquad
  A^-_\nu A^+_\nu=L_{\nu}^{iso}-2\nu+1\,,&\\
&  A^-_\nu L_{\nu-1}^{iso}=(L_{\nu}^{iso}+2)A^-_\nu\,,\qquad
    A^+_\nu L_{\nu}^{iso}=(L_{\nu-1}^{iso}-2)A^+_\nu\,.&\label{AL+A}
 \end{eqnarray}
 Unlike the case of the harmonic oscillator,
 the partner Hamiltonian operators are described 
 by potentials  characterized by different coupling constants,
 and though we have a shape invariance  \cite{shape1,shape2},
 the first order differential operators $A^+_\nu$
 and $A^-_\nu$ are not ladder operators
 for $L_{\nu}^{iso}$ or $L_{\nu-1}^{iso}$.
 One can use the 
symmetry of the coupling constant  
 and consider the first order operators 
 by changing $\nu\rightarrow -\nu-1$ 
 in (\ref{A1nu-def}).
 The obtained in such a way operators 
 will factorize and intertwine
 the pair of the Hamiltonian operators 
  $L_{\nu}^{iso}$  and 
$L_{\nu+1}^{iso}$.
 We then make additionally a shift $\nu\rightarrow \nu-1$, and 
 obtain the first order differential operators
 \be\label{A-nudef}
 A^-_{-\nu}=-A^+_{\nu}{}_{\vert}{}_{\nu\rightarrow -\nu}=
 \frac{d}{dx}-x-\frac{\nu}{x}\,,\qquad
 A^+_{-\nu}=-A^-_{\nu }{}_{\vert}{}_{\nu\rightarrow -\nu}=
- \frac{d}{dx}-x-\frac{\nu}{x}\,,
\ee
for which 
$\text{ker}\,A^+_{-\nu}=\psi_{-\nu-1,0}=x^{-\nu}e^{-x^2/2}$
and $\text{ker}\,A^-_{-\nu}=(\psi_{-\nu-1,0})^{-1}$.
It may be noted  that $(\psi_{-\nu-1,0}(x))^{-1}=\psi_{\nu-1}(ix)$
is a non-physical eigenstate of $L_{\nu-1}^{iso}$ of eigenvalue
$-2\nu-3=-E_{\nu-1,0}$. Operators (\ref{A-nudef}) satisfy the relations 
  \begin{eqnarray}\label{AA-nu}
 &A^+_{-\nu} A^-_{-\nu}=L_{\nu-1}^{iso}+2\nu+1\,,\qquad
  A^-_{-\nu} A^+_{-\nu}=L_{\nu}^{iso}+2\nu-1\,,&\\
&  A^-_{-\nu} L_{\nu-1}^{iso}=(L_{\nu}^{iso}-2)A^-_{-\nu}\,,\qquad
    A^+_{-\nu} L_{\nu}^{iso}=(L_{\nu-1}^{iso}+2)A^+_{-\nu}\,.&\label{AL-A}
 \end{eqnarray}
In  
 (\ref{AA-nu}) the  additive factorization constants 
are opposite in sign to those in  (\ref{AA+nu}),
and the same is true for the displacement  constants 
in intertwining relations (\ref{AL-A}) and (\ref{AL+A}). 
Combining intertwining relations (\ref{AL-A}) and (\ref{AL+A}),
one can construct  ladder operators for 
the system (\ref{isoham}),
\begin{eqnarray}\label{Cnu+-1}
&\mathfrak{C}_{\nu}^-=A^-_\nu A^+_{-\nu}=A^+_{-\nu-1}A^-_{\nu+1}=-(a^-)^2-\frac{\nu(\nu+1)}{x^2}\,,&\\
&\mathfrak{C}_{\nu}^+=A^-_{-\nu} A^+_\nu=A^+_{\nu+1}A^-_{-\nu-1}=-(a^+)^2-\frac{\nu(\nu+1)}{x^2}\,.&
\label{Cnu+-2}
\end{eqnarray}
They satisfy relations 
$[L^{iso}_\nu,\mathfrak{C}_{\nu}^\pm]=\pm 4\mathfrak{C}_{\nu}^\pm$
and $\mathfrak{C}_{\nu}^\pm\psi_{\nu,n}\propto \psi_{\nu,n\pm 1}$, 
$\mathfrak{C}_{\nu}^-\psi_{\nu,0}=0$
in correspondence with Eq. (\ref{Enun}).
\vskip0.1cm

The above comments on the choice of the seed states 
in the form of  non-physical eigenstates 
obtained by Wick rotation $x\rightarrow ix$ from  the 
ground states of the harmonic and isotonic oscillators
correspond to a particular case of the already noted
 specific symmetry 
of both quantum systems.
In correspondence with this,  
functions $ \psi_n(ix)\equiv \psi_{-n}(x)$ 
and $\psi_{\nu,n}(ix)\equiv \psi_{\nu,-n}(x) $ are  formal (non-physical)
eigenstates of $L_0$ and $L^{iso}_\nu$ of eigenvalues $\lambda=-(2n+1)$
and $\lambda=-(2\nu+4n+3)$, respectively,
which are non-normalizable, exponentially increasing at infinity 
functions. 
Regardless of their non-physical nature, 
 these states also  can be used as seed states for  DCKA  transformations
 and  will play a key role in our further constructions.
 Furthermore, we have 
solutions 
(\ref{secondlinearindependent}), which 
by means of relation (\ref{Darbouxstates}), or equivalently, 
$\psi_{[n,\lambda]}=\A_n\psi_\lambda$,
 may be transformed into normalizable wave functions. 
When this happens,
the transformed  system has a 
corresponding  eigenstate 
 in the lower part of the spectrum.

\vskip0.1cm

For the sake of  brevity
we will use the following notations for physical and non-physical
eigenstates of the QHO\,:
\begin{equation}
\label{notation}
n\equiv \psi_n(x),\qquad -n\equiv \psi_{-n}=\psi_n(ix)\,,\qquad \widetilde{n}\equiv \widetilde{\psi_n}\,,
\qquad \widetilde{-n}\equiv \widetilde{\psi_{-n}}.
\end{equation}  
Each rational deformation of the QHO and  AFF systems 
we will consider below can be generated  by different  DCKA transformations 
from the  harmonic and the half-harmonic oscillators.
When all the seed states
are eigenstates with positive 
index $n>0$, following \cite{CIP} we call the corresponding DCKA scheme positive.
When the seed states carry only negative integer index of
non-physical eigenstates   
and the DCKA transformation  produces the same 
system as the positive scheme modulo the global shift of the spectrum,
we call such a scheme negative, and refer to the 
``complementary"  schemes as dual.
For the details on the origin of such a duality see \cite{CIP}.

\subsection{$\mathfrak{osp}(2|2)$ and super-Schr\"odinger symmetries}\label{Section3.2}

We start with the $\mathcal{N}=2$ superextended  AFF 
model described by the matrix Hamiltonian operator
\be
\label{hamiliso}
\mathcal{H}_\nu=
\left(
\begin{array}{cc}
L_{\nu}^{iso}-2\nu+1&    0 \\
0 &  L_{\nu-1}^{iso}-2\nu-1    
\end{array}
\right).
\ee
With this definition, the equidistant spectrum of the system 
is given by $\mathcal{E}_n=4n,$  $n=0,1,\ldots$,
where $n=0$ corresponds to the nondegenerate ground state 
$(0,\psi_{\nu-1,0})^t$ of zero energy, 
while  all energy levels with $n\geq 1$ are doubly degenerate.
The described  properties of the AFF model 
allow us to identify the Lie-superalgebraic symmetry  of 
the extended system (\ref{hamiliso}) generated 
by the even,
$\mathcal{H}_\nu$, $\mathcal{R}_\nu=\Sigma-\nu \mathbb{I}$, $\mathcal{C}_\nu^\pm$,
and odd, 
$\mathcal{Q}_\nu^a$, 
$\mathcal{S}_\nu^a$, operators, 
where $\mathbb{I}$ is the unit $2\times 2$ matrix, and 
\begin{eqnarray}\label{lader}
&\mathcal{C}_\nu^{\pm}=
\left(
\begin{array}{cc}
\mathfrak{C}_{\nu}^\pm&   0  \\
 0 &  \mathfrak{C}_{\nu-1} ^\pm    
\end{array}
\right),&\\
\label{supercharge1}
&\mathcal{Q}_{\nu}^{1}=
\left(
\begin{array}{cc}
  0&    A^-_\nu  \\
 A^+_\nu &   0     
\end{array}
\right),
\qquad
 \mathcal{S}_\nu^1=
\left(
\begin{array}{cc}
  0&    A^-_{-\nu}  \\
 A^+_{-\nu} &   0     
\end{array}
\right),&\\
&\mathcal{Q}_\nu^{2}=
i\sigma_3\mathcal{Q}_\nu^1\,,
\qquad \mathcal{S}_\nu^2=
i\sigma_3\mathcal{S}_\nu^1\,.&
\end{eqnarray}
Here the bosonic generators $\mathcal{C}_\nu^{\pm}$ are composed 
from the ladder operators 
given by Eqs. (\ref{Cnu+-1}),
(\ref{Cnu+-2}) 
and their analogs with index $\nu$ changed to $\nu-1$.
The supercharges $\mathcal{Q}_\nu^a$ 
correspond to the odd integrals defined in (\ref{Hlambda*}), and 
the  odd generators $\mathcal{S}_\nu^a$
are constructed from the intertwining 
operators (\ref{A-nudef}).
The Lie superalgebraic relations have the form 
\begin{eqnarray}\label{HRQ0}
&[\mathcal{H}_\nu,\mathcal{R}_\nu]=[\mathcal{H}_\nu,\mathcal{Q}_\nu^a]=0\,,&\\
\label{evencommutation}
&[\mathcal{H}_\nu,\mathcal{C}_\nu^{\pm}]=\pm4\mathcal{C}_\nu^{\pm}\,, \qquad 
[\mathcal{C}_\nu^{-},\mathcal{C}_\nu^{+}]=8\mathcal{H}_\nu-16\mathcal{R}_\nu\,,&\\
\label{evenodd}
&[\mathcal{H}_\nu,\mathcal{S}_\nu^a]=-4i\epsilon^{ab}\mathcal{S}_\nu^b\,,\qquad
[\mathcal{R}_\nu,\mathcal{Q}_\nu^a]=-i\epsilon^{ab}\mathcal{Q}_\nu^b\,,
\qquad
[\mathcal{R}_\nu,\mathcal{S}_\nu^a]=-i\epsilon^{ab}\mathcal{S}^b_\nu\,,&\\
\label{fq1}
&[\mathcal{C}_\nu^-,\mathcal{Q}_\nu^a]=2(\mathcal{S}_\nu^a+i\epsilon^{ab}\mathcal{S}_\nu^b), \qquad 
[\mathcal{C}_\nu^+,\mathcal{Q}_\nu^a]=-2(\mathcal{S}_\nu^a-i\epsilon^{ab}\mathcal{S}_\nu^b)\,,&\\
\label{fq3}
&[\mathcal{C}_\nu^-,\mathcal{S}_\nu^a]=2(\mathcal{Q}_\nu^a-i\epsilon^{ab}\mathcal{Q}_\nu^b)\,, \qquad 
[\mathcal{C}_\nu^+,\mathcal{S}_\nu^a]=-2(\mathcal{Q}_\nu^a+i\epsilon^{ab}\mathcal{Q}_\nu^b)\,,&\\
\label{anti1}
&\{ \mathcal{Q}_\nu^a,\mathcal{Q}_\nu^b\}=2\delta^{ab}\mathcal{H}_\nu\,, \qquad 
\{ \mathcal{S}_\nu^a,\mathcal{S}_\nu^b\}=2\delta^{ab}(\mathcal{H}_\nu -4\mathcal{R}_\nu)\,,&\\
\label{anti2}
&\{\mathcal{Q}^a_\nu,\mathcal{S}^b_\nu\}=\delta^{ab}(\mathcal{C}_\nu^{+}+\mathcal{C}_\nu^-)+
i\epsilon^{ab}(\mathcal{C}_\nu^+-\mathcal{C}_\nu^-)\,,&
\end{eqnarray}
and correspond to superconformal $\mathfrak{osp}(2|2)$
symmetry.  Since both supercharges $\mathcal{Q}_\nu^a$
annihilate the unique ground state, 
the $\mathcal{N}=2$ super-Poincar\'e symmetry 
is unbroken.
The noncommutativity of $\mathcal{S}_\nu^a$  
with $\mathcal{H}_\nu$ appears because 
the operators $A^\pm_{-\nu}$ 
intertwine $L^{iso}_\nu$ and $L^{iso}_{\nu-1}$
with different  shift constants in comparison 
with  $A^\pm_{\nu}$. Nonzero commutators
of $\mathcal{C}_\nu^\pm$ with $\mathcal{H}_\nu$
correspond to the commutation relations 
of the respective ladder operators  with  Hamiltonians of the super-partner
subsystems.
So, the generators $\mathcal{S}_\nu^a$ and $\mathcal{C}_\nu^\pm$
are not integrals of motion
of the system $\mathcal{H}_\nu$. 
They can be promoted to the \emph{dynamical integrals} 
of motion via time-dressing by the evolution
operator,
\begin{eqnarray}
\label{recipeintegrals}
&\mathcal{O}_\nu\rightarrow \widetilde{\mathcal{O}_\nu}= e^{-i\mathcal{H}_\nu t}\mathcal{O}_\nu 
e^{i\mathcal{H}_\nu  t}\,.&
\end{eqnarray} 
The transformed in this way operators 
are explicitly depending on time 
integrals of motion 
$\widetilde{\mathcal{S}_\nu^1}=\mathcal{S}_\nu^1\cos 4t -\mathcal{S}_\nu^2\sin4t$,
$\widetilde{\mathcal{S}_\nu^2}=\mathcal{S}_\nu^2\cos 4t +\mathcal{S}_\nu^1\sin4t$,
and $\widetilde{\mathcal{C}_\nu^\pm}=e^{\mp 4it}\mathcal{C}_\nu^\pm$,
which satisfy  the Heisenberg equation of
motion of the form  $\frac{d}{dt}\widetilde{\mathcal{O}_\nu}=\partial \widetilde{\mathcal{O}_\nu}/\partial t-i
[\widetilde{\mathcal{O}_\nu},\mathcal{H}_\nu]=0$.
The substitution of  $\mathcal{S}_\nu^a$  and  $\mathcal{C}_\nu^\pm$
for the time-dressed integrals $\widetilde{\mathcal{S}_\nu^a}$ and 
$\widetilde{\mathcal{C}_\nu^\pm}$ does not change the form of
the  $\mathfrak{osp}(2|2)$ superalgebraic relations 
since transformation (\ref{recipeintegrals}) is unitary. 

\vskip0.1cm

We have fixed  the Hamiltonian operator (\ref{hamiliso}) 
coherently with  the supercharges $\mathcal{Q}_\nu^a$ 
constructed in terms of the intertwining  operators $A^\pm_\nu$.
The matrix operators  $\mathcal{S}_\nu^a$ constructed 
in terms of the other pair of  intertwining  operators $A^\pm_{-\nu}$
turn out in this case to be  the dynamical integrals of motion.
If 
we choose, instead, the operator 
 $\mathcal{H}'=\mathcal{H}_\nu -4\mathcal{R}_\nu=\text{diag}\,(L^{iso}_\nu+2\nu-1,\, 
 L^{iso}_{\nu-1}+2\nu+1) $
 as a Hamiltonian of the superextended system,
then fermionic operators $\mathcal{S}_\nu^a$
will be true, not depending explicitly on time, integrals of motion (supercharges),
while $\mathcal{Q}_\nu^a$ will transform (after unitary time-dressing procedure)
 into dynamical integrals of motion.
In effect, these changes are a part  of the transformation 
$\mathcal{H}_\nu\rightarrow \mathcal{H}'_\nu=\mathcal{H}_\nu -4\mathcal{R}_\nu$, 
$\mathcal{Q}_\nu^1 \rightarrow {\mathcal{Q}'}_\nu^1=\mathcal{S}_\nu^2$,
$\mathcal{Q}_\nu^2 \rightarrow {\mathcal{Q}'}_\nu^2=\mathcal{S}_\nu^1$,
$\mathcal{S}_\nu^1 \rightarrow {\mathcal{S}'}_\nu^1=\mathcal{Q}_\nu^2$,
$\mathcal{S}_\nu^2 \rightarrow {\mathcal{S}'}_\nu^2=\mathcal{Q}_\nu^1$
$\mathcal{R}_\nu \rightarrow  \mathcal{R}'_\nu=-\mathcal{R}_\nu$,
$\mathcal{C}_\nu^\pm \rightarrow  
{\mathcal{C}'}_\nu^\pm= \mathcal{C}_\nu^\pm$,
which  is the automorphism of the superalgebra 
(\ref{HRQ0})--(\ref{anti2}).
The superextended system described by the 
Hamiltonian $\mathcal{H}'_\nu=\mathcal{H}_\nu -4\mathcal{R}_\nu$
has, however, essentially different properties in comparison with the
system given by the matrix Hamiltonian $\mathcal{H}_\nu$.
Its spectrum is $\mathcal{E}_n=4n+2\nu+2$, $n=0,1,\ldots$,
and all its energy levels including the lowest one, 
$\mathcal{E}_0=+2\nu+2>0$, are doubly degenerate, 
and neigther of its two lowest eigenstates is 
annihilated  by both supercharges ${\mathcal{Q}'}_\nu^a$.
This means that the system $\mathcal{H}'_\nu$, unlike 
$\mathcal{H}_\nu$, is characterized by the 
spontaneously broken $\mathcal{N}=2$ 
super-Poincar\'e symmetry. 
 
 \vskip0.1cm

One  can change the basis by considering the  linear combinations
\begin{eqnarray}
&\pi_{\nu}^a=\frac{1}{2\sqrt{2}}(\mathcal{S}_\nu^a -\epsilon^{ab}\mathcal{Q}_\nu^b)\, ,
\qquad 
\zeta_\nu^{a}=\frac{1}{2\sqrt{2}}(\mathcal{Q}_\nu^a-\epsilon^{ab}\mathcal{S}_\nu^b)\,, &\\
&\mathcal{D}_\nu=\frac{1}{8}(\mathcal{C}_\nu^++\mathcal{C}_\nu^-)\,,&\\
&\mathcal{L}_\nu=\frac{1}{4}\mathcal{H}_\nu-\frac{1}{2}\mathcal{R}_\nu+
\frac{i}{2}(\mathcal{C}_\nu^+-\mathcal{C}_\nu^-)\,,\qquad
\mathcal{K}_\nu=\frac{1}{4}\mathcal{H}_\nu-\frac{1}{2}\mathcal{R}_\nu-
\frac{i}{2}(\mathcal{C}_\nu^+-\mathcal{C}_\nu^-)\,. &
\end{eqnarray}
Then nonzero (anti)commutation relations 
of the superalgebra  take   the form
 \begin{eqnarray}
 &[\mathcal{D}_\nu,\mathcal{L}_\nu]=i\mathcal{L}_\nu\,,\qquad 
 [\mathcal{D}_\nu,\mathcal{K}_\nu]=-i\mathcal{K}_\nu\,,\qquad 
 [\mathcal{K}_\nu,\mathcal{L}_\nu]=2i\mathcal{D}_\nu\,,&\\
&\{\pi_\nu^{a},\pi_\nu^{b}\}=2\delta^{ab}\mathcal{L}_\nu\,,
\qquad
\{\zeta_\nu^{a},\zeta_\nu^{b}\}=2\delta^{ab}\mathcal{K}_\nu\,,
\qquad \{\pi_\nu^{a},\zeta_\nu^{b}\}=2\delta^{ab}\mathcal{D}_{\nu}+\epsilon^{ab}\mathcal{R}_\nu\,,&\\
&[\mathcal{L}_\nu,\zeta_\nu^{a}]=-i\pi_\nu^{a}\,,\qquad [\mathcal{K}_\nu,\pi_\nu^{a}]=i\zeta_\nu^{a}\,,\qquad
[\mathcal{D}_\nu,\pi_\nu^{a}]=\frac{i}{2}\pi_\nu^{a}\,,\qquad
[\mathcal{D}_\nu,\zeta_\nu^{a}]=-\frac{i}{2}\zeta_\nu^{a}\,,&\\
&[\mathcal{R}_\nu,\pi_\nu^{a}]=i\epsilon^{ab}\pi_\nu^{b}\,,\qquad 
[\mathcal{R}_\nu,\zeta_\nu^{a}]=i\epsilon^{ab}\zeta_\nu^{b}\,.&
\end{eqnarray}
This is a form of superconformal algebra which appears in a free 
nonrelativisitc spin-$1/2$ particle system and 
superextended  AFF model without the confining harmonic  potential term \cite{AnaPLy}.
One can note  that 
if we restore the harmonic oscillator frequency $\omega$, which was fixed here equal to $2$,  
and  take its zero limit, 
we recover the supersymmetric two-particle Calogero model 
(with the omitted center of mass degree of freedom)
and its superconformal symmetry.

\vskip0.1cm 

The superconformal symmetry of the superextended QHO 
described by matrix Hamiltonian $\mathcal{H}=\text{diag}\,(L_0+1,L_0-1)$
can be generated in analogous way by  taking $\psi_0$ and $\psi_{-0}$ 
as the seed states to produce ladder operators as the intertwining-factorizing 
operators in two different ways as it was described above.
The superconformal structure of the supersymmetric harmonic oscillator
can be obtained, however, in a more direct way  just by \emph{formally} putting 
$\nu=0$ in the Hamiltonian operator (\ref{isoham}), in the first order intertwining operators (\ref{A1nu-def})
and (\ref{A-nudef}), in the ladder operators 
(\ref{Cnu+-1}), (\ref{Cnu+-2}) of the AFF model, and in  the generator $\mathcal{R}_\nu$ 
of the $U(1)$ $\mathcal{R}$-symmetry.
In such a way we reproduce  the superconformal $\mathfrak{osp}(2|2)$
algebra for the supersymmetric QHO
having exactly  the same form as  for the superextended AFF system.
The important point, however,  is that the QHO 
is defined on the  whole  real line instead of the half-line in the case of the
AFF model. Because of this the distance between its energy levels is twice 
less than in the AFF model and the superconformal $\mathfrak{osp}(2|2)$
symmetry expands upto the superextended Schr\"odinger symmetry
whose superalgebra includes the superextended Heisenberg 
algebra in the form of  invariant subsuperalgebra.

The operators $\mathcal{C}_0^\pm=-(a^\pm)^2\equiv  \mathcal{J}_\pm$,  unlike the case of the 
superextended 
AFF model (\ref{hamiliso}),  are not enough now 
to produce all the Hilbert space of the super-oscillator if we start from its ground state 
$(0,\psi_0)^t$ 
and also use other generators of the $\mathfrak{osp}(2|2)$\,:
in this way we generate only the states of the form
$(0,\psi_{2n})^t$ and $(\psi_{2n+1},0)^t$, $n=0,1,\ldots$,
and their linear combinations.
To produce the missing states, it is sufficient to extend the set of
generators by matrices $\Sigma_a=\sigma_a$, $a=1,2$.
They anti-commute with the grading operator $\Gamma=\sigma_3$,
and are identified as odd generators. {}From the commutation 
relations $[\mathcal{H},\Sigma_\pm]=\pm 2\Sigma_\pm$,
where $\Sigma_\pm=\frac{1}{2}(\Sigma_1\pm i\Sigma_2)$,
we find that by unitary transformation of the form (\ref{recipeintegrals})
they can be promoted to the dynamical integrals 
$\widetilde{\Sigma_\pm}=e^{\mp 2it}\Sigma_\pm$.
The anti-commutators of $\Sigma_a$ between themselves 
generate central charge $\mathbb{I}$, while their anti-commutators  with 
fermionic generators $\mathcal{Q}_a$ and $\mathcal{S}_a$
produce a new pair of the bosonic generators 
\begin{equation}
\mathcal{G}_{\pm}=
\left(
\begin{array}{cc}
a^\pm & 0    \\
0 &  a^\pm     
\end{array}
\right).
\end{equation}
Operators $\mathcal{G}_\pm$ together with 
the identity matrix operator $\mathbb{I}$
generate the $\mathfrak{h}_1$ Heisenberg  algebra
which is an invariant subalgebra of the 
complete superextended Schr\"odinger symmetry 
of the super-oscillator.  Operators
$\mathcal{G}_\pm$, $\mathbb{I}$ and $\Sigma_a$ 
generate the superextended 
Heisenberg  algebra.
The complete set of the (anti)-commutation relations 
of the superextended Schr\"odinger symmetry 
is given by those of the $\mathfrak{osp}(2|2)$ 
superalgebra and  by  relations 
\begin{eqnarray}
\label{qho3}
&[\mathcal{H},\mathcal{G}_\pm]=\pm2\mathcal{G}_{\pm}\,,\qquad
[\mathcal{G}_\mp,\mathcal{J}_{\pm}]=\mp 2\mathcal{G}_{\pm}\,,\qquad
[\mathcal{G}_-,\mathcal{G}_{+}]=2\mathbb{I}\,,&\\
&\{\Sigma_{a},\Sigma_b\}=2\delta_{ab}\mathbb{I}\,,\qquad [\mathcal{H},\Sigma_a]=
2i\epsilon_{ab}\Sigma_{b} \,,\qquad [\sigma_3,\Sigma_a]=-2i\epsilon_{ab}\Sigma_{b}\,,&\\
&\{\Sigma_{a},\mathcal{Q}_b\}=\delta_{ab}(\mathcal{G}_{+}+\mathcal{G}_{-})+
i\epsilon_{ab}(\mathcal{G}_{+}-\mathcal{G}_{-})\,,&\\
&\{\Sigma_{a},\mathcal{S}_b\}=\delta_{ab}(\mathcal{G}_{+}+\mathcal{G}_{-})-
i\epsilon_{ab}(\mathcal{G}_{+}-\mathcal{G}_{-})\,,&\\
&[\mathcal{G}_-,\mathcal{Q}_a]=\Sigma_a+i\epsilon_{ab}\Sigma_b\, ,\qquad 
[\mathcal{G}_+,\mathcal{Q}_a]=-\Sigma_a+i\epsilon_{ab}\Sigma_b\,,&\\
\label{qho4}
&[\mathcal{G}_-,\mathcal{S}_a]=\Sigma_a-i\epsilon_{ab}\Sigma_b \,,\qquad 
[\mathcal{G}_+,\mathcal{S}_a]=-(\Sigma_a+i\epsilon_{ab}\Sigma_b)\,,&\\
\label{SigGJ0}
&[\Sigma_{a},\mathcal{G}_\pm]=[\Sigma_{a},\mathcal{J}_\pm]=0\,.&
\end{eqnarray}
Let us also note that generators $\Sigma_a$ 
 intertwine the supercharges 
$\mathcal{Q}_a$ with the dynamical fermionic 
integrals $\mathcal{S}_a$\,:
$\Sigma_1\mathcal{Q}_a=
\epsilon_{ab}\mathcal{S}_b\Sigma_1$,
$\Sigma_2\mathcal{Q}_a=-\epsilon_{ab}\mathcal{S}_b\Sigma_2$.
It is also worth to note that  if we try to expand  the set of symmetry generators
of the superextended AFF system by $\Sigma_a$, their 
(anti)-commutation with generators  of the $\mathfrak{osp}(2|2)$ symmetry
would lead to  infinite-dimensional superalgebraic structure.
In other way this can be understood from the point of view of the 
time-dressing procedure. In this case 
$[\mathcal{H}_\nu,\Sigma_{a}]\sim 1/x^2\cdot \epsilon_{ab}\Sigma_b$,
and the unitary transformation (\ref{recipeintegrals}) results 
in a nonlocal operator being an infinite series in derivative 
$d/dx$ and time parameter $t$.

\vskip0.1cm
If similarly to the case of the superextended AFF model,
we change the Hamiltonian $\mathcal{H}_0$ to 
$\mathcal{H}_0'=\mathcal{H}_0-4\mathcal{R}_0=\text{diag}\,(L_0-1,L_0+1)$,
we obtain  exactly the same matrix operator but with just
permuted Hamiltonian operators of the subsystems. 
This transformation is a part  of the corresponding automorphism of 
the super-Schr\"odinger algebra which, unlike the AFF model case, 
does not change the unbroken nature  of the $\mathcal{N}=2$ super-Poincar\'e symmetry.
\vskip0.1cm

In the limit of zero frequency, the  super-Schr\"odinger 
symmetry of the  free  nonrelativistic spin-$1/2$  quantum particle is recovered 
\cite{BeckHussin,DuvHorSh,InzPly}.   
This  also is coherent with the known relation between the 
free particle system, harmonic oscillator, and AFF model \cite{AFF,BTGE,Nied}.
If  we change the coordinate and time variable of the free particle
of mass $M$ as 
 $(x,t)\rightarrow (\xi,\tau)$,
$x(\xi,\tau)=\xi(\beta\omega)^{1/2}/\cos (\omega\tau)$, $t(\tau)= \beta\tan (\omega\tau)$,
where $\beta$ is a real constant of dimension of time,
its action $\mathcal{A}=\frac{1}{2}M\int\dot{x}^2dt$ will  be transformed into
the action $\mathcal{A}_\omega=\int L(\xi,\xi')d\tau$ of the harmonic oscillator 
of frequency $\omega$
with 
\begin{eqnarray}
&L(\xi,\xi')=\frac{1}{2}M\left((\xi')^2-\omega^2\xi^2\right)+\frac{d}{d\tau}\left(\frac{1}{2}M\omega
\xi^2\tan(\omega\tau)\right)\,,&
\end{eqnarray}
 where $\xi'=d\xi/d\tau$.The total derivative term is essential for getting 
 the QHO's  propagator from the free particle system's  propagator
\cite{OleS,BTGE}.
In the context of (super)conformal symmetry we discuss,
 it is important that
the   
action potential term $\mathcal{A}_g=g\int \frac{dt}{x^2}$
is invariant under the indicated transformation,
 $\mathcal{A}_g\rightarrow g\int \frac{d\tau}{\xi^2}$.
 Then  the same transformation applied to the 
 Calogero system given by Lagrangian $L=\frac{1}{2}M\dot{x}^2+\frac{g}{x^2}$
 will produce the Lagrangian for the AFF model with 
 the confining harmonic oscillator potential term.
 This is  a particular case 
of the so called classical Arnold 
 transformation \cite{AT}.  
\vskip0.1cm

For the superconformal symmetries of the 
$\mathcal{N}=2$ superextended  AFF and QHO systems
there  was important the equidistant nature of their spectra,
and that the sets of the spectrum generating operators of the initial  
quantum models  are composed by the conjugate pairs of 
differential operators of the second and first order, respectively.
The rational deformations of these systems are characterized  
in a generic case by the sets of three pairs of the spectrum generating 
ladder operators which are differential operators of the orders not lower than 
three. This difference, as we shall see, leads to the 
essentially different \emph{nonlinear} structures of the superconformal 
and super-Schr\"odinger symmetries of the corresponding 
$\mathcal{N}=2$ superextended systems.
\vskip0.1cm

To conclude this section we note that as it follows from relations  
(\ref{HRQ0})--(\ref{anti2}), the complete  $\mathfrak{osp}(2|2)$ superconformal structure
of the superextended AFF  system
can be obtained by extending the set of  its generators 
of $\mathcal{N}=2$ supersymmetry  (\ref{N2susy})
by any one of the dynamical integrals  $\mathcal{S}^a_\nu$,  $\mathcal{C}^+_\nu$ 
or  $\mathcal{C}^-_\nu$ (or by any of the operators $\mathcal{S}_a$, $\mathcal{J}_+$ or
$\mathcal{J}_-$ in the case of super-oscillator).
To recover  additional (anti)-commutation relations 
(\ref{qho3})--(\ref{anti2})
of the super-Schr\"odinger symmetry of super-oscillator system,
it is sufficient then to extend the set of 
generators of its $\mathfrak{osp}(2|2)$ superalgebra
by any one of the dynamical integrals 
$\Sigma_a$, $a=1,2$, $\mathcal{G}_+$ or $\mathcal{G}_-$.
This observation will play a key role 
in what follows for generation of nonlinearly deformed and 
extended superconformal and super-Schr\"odinger 
symmetries of the $\mathcal{N}=2$ super-extensions 
of the rationally deformed  AFF and harmonic oscillator systems.

\section{Symmetry generators}
\label{gen}

We pass now to  the study  of  extensions and deformations 
of the superconformal and super-Schr\"odinger symmetries 
which appear  in the $\mathcal{N}=2$ superextended systems 
described by the pairs of the Hamiltonian operators 
($L_0$, $L_{0,def}$) and ($L_{0^+}$,  $L^{iso}_{m,def}$).
Here $L_{0,def}$ and    $L^{iso}_{m,def}$ correspond
to rational deformations of the QHO
and the AFF model with integer 
values of the parameter $\nu=m$, $m\in \N$. 
In general case, rationally
deformed  harmonic and isotonic oscillator  systems
are characterized by finite number of gaps (missing energy levels) 
in their spectra, and their description requires 
more than one pair of spectrum generating ladder operators.
It is because of this  expansion of the sets of ladder  operators
and of their higher  differential order that
there appear  
nonlinearly deformed superconformal 
and super-Schr\"odinger structures which generalize  
the corresponding Lie superalgebraic symmetries.
This section is devoted to the description  of  the complete sets of 
generators of the indicated symmetries.
They will be constructed  on the basis of 
the QHO.

\subsection{Generation of rationally extended oscillator systems}

We first consider rational deformations (extensions) 
 of the harmonic oscillator.
They are constructed following the well known recipe of the 
Krein-Adler theorem \cite{Krein,Adler}. 
Consider a  `positive' scheme  $(n_1,n_1+1,\ldots,n_\ell,n_\ell+1)$, 
where  $n_j\in \N $, $j=1,\ldots,\ell$, correspond to  the 
set of the chosen seed states, see notations  (\ref{notation}).
It generates the deformed system 
\begin{eqnarray}
\label{DQHO}
&L_{(n_1,n_1+1,\ldots,n_\ell,n_\ell+1)}=L_{0} +4\ell +
\frac{F(x)}{Q(x)},&
\end{eqnarray} 
where $F(x)$ and $Q(x)$ are real-valued even polynomials,
with $Q(x)$ being positive definite and having the degree of $F(x)$ plus two.
Physical eigenstates of this system are obtained by the map
(\ref{Darbouxstates}) applied to the physical eigenstates
of harmonic oscillator with indexes different from those of the seed states.
The resulting system has certain number of gaps
in the spectrum at 
 location of energy levels of the seed states,
and each gap corresponds to even number of missing  neighbour 
energy levels.
\vskip0.1cm

Another class of systems which can be constructed by 
Darboux-Crum transformations on the basis
of the QHO
corresponds to deformations of the AFF model
(\ref{isoham}) with $\nu=m$, $m\in \N$.
For this one can first construct rationally deformed harmonic oscillator system
as in (\ref{DQHO}), then introduce an infinite potential barrier at $x=0$, 
and finally take first $m$ physical states (satisfying Dirichlet boundary condition at $x=0$)
as the seed states for additional Darboux-Crum transformation.
This composite generating  procedure admits other interpretations due to 
the iterative properties of DCKA transformations.
One way to see this is to take the half-harmonic oscillator   
$L_{0^+}$ as a starting system and use 
the set of states  $(n_1,n_1+1,\ldots,n_\ell,n_\ell+1,2k_1+1,\ldots, 2k_m+1)$, 
where even indexes inside the set  
$n_1,n_1+1,\ldots,n_\ell,n_\ell+1$ represent non-physical eigenstates 
of $L_{0^+}$ and 
$k_i$, $i=1,\ldots,m$,   are identified as  
$m$ odd states which were not considered in the first set  of $2n_\ell$ states.
The  Hamiltonian operator
\begin{eqnarray}
\label{REIO}
&L_{(n_1,n_1+1,\ldots,n_
\ell,n_\ell+1,2k_1+1,\ldots, 2k_m+1)}=L_{m}^{iso}+
2m +4\ell + \frac{\widetilde{F(x)}}{\widetilde{Q(x)}},&
\end{eqnarray} 
appears as a final result of the described procedure,
where polynomials $\widetilde{F(x)}$ and  $\widetilde{Q(x)}$  
have the properties similar to those of $F(x)$ and $Q(x)$ in 
(\ref{DQHO}).
With the described method one can obtain 
rational deformations of the AFF model with 
integer values of index $\nu=m$ only,  on which we will focus  here.
Physical states of  the  system (\ref{REIO})  are obtained 
by applying the mapping 
(\ref{Darbouxstates})  to \emph{odd} states of harmonic oscillator 
with index different from that of the seed states. In general such a 
system has gaps in its spectrum. 
If, however,  the set
$n_1,n_1+1,\ldots,n_
\ell,n_\ell+1,2k_1+1,\ldots, 2k_m+1$ contains all
the $\ell+m$ odd indexes from $1$ to $2k_m+1$,
the generated deformed AFF system 
will have  no gaps in its spectrum  
and  we obtain a system completely isospectral
to $L_{0^+}+4\ell+2m$. Such completely isospectral
(gapless)  rational deformations are  impossible 
in the case of the harmonic oscillator system.
\vskip0.1cm

The method of mirror  diagrams developed and employed in \cite{CIP} is a
technique 
 such that a dual scheme with
non-physical  `negative'  eigenstates  (\ref{notation}),  which  
are
 obtained by the
transformation $x\rightarrow ix$ from physical states,
is derived from a positive scheme with physical states of $L_0$ only,
and vice versa.
Both schemes are related in such a way that 
the second  logarithmic derivatives
of their Wronskians are equal up to an additive  constant.
Formally one  scheme can be obtained from another in the following way.
Let positive scheme is 
$\Delta^+\equiv (l_1^+,\ldots,l_{n_+}^+)$, where 
$l_i^+$ with  $i=1,\ldots, n_+$ are certain positive numbers 
which have to be chosen according to the
already described rules. 
Then    negative scheme is given by 
 $\Delta^-=(-\check{0},\ldots,-\check{n}_i^-,
\ldots,-l_{n_+}^+)$, where $-\check{n}_i^-\equiv l_i^+-l_{n_+}^+$,  
means that the corresponding number $-n_i^-\equiv l_i^+-l_{n_+}^+$ is omitted
from  the set  $\Delta^-$.
On the contrary, if we are given  the negative scheme 
$\Delta^-\equiv (-l_{1}^-,\ldots,-l_{n_-}^-)$,  where   $-l_{j}^-$  with 
$j=1,\ldots, n_-$ 
are certain negative numbers, 
then positive scheme is given by 
 $\Delta^+=(\check{0}, \ldots,\check{n}^+_{j},\ldots,l_{n_-}^-)$, where symbols 
$\check{n}_j^+=l_{n_-}^- -l_{j}^-$ represent the 
states missing from  the list of the chosen seed states.  
For dual schemes the relation 
\be\label{genDual}
e^{-n_+x^2/2}W(-l_{1}^-,\ldots,-l_{n_-}^-)=e^{n_-x^2/2}
W(l_1^+,\ldots,l_{n_+}^+)\,,
\ee
is valid modulo a multiplicative constant, from which one finds that
the Hamiltonians of dual schemes satisfy the relation 
\begin{equation}
\label{L+L-}
L_{(+)}-L_{(-)}=2N\,,\qquad N\equiv n_+ +n_-=l_{n^+}^++1=l_{n^-}^-+1\,,
\end{equation}
where  $L_{(+)}\equiv  L_{(l_{1}^+,\ldots,l_{n_+}^+)}$ and
 $ L_{(-)}\equiv  L_{(-l_{1}^-,\ldots,-l_{n_-}^-)}$.

By means of negative scheme we do not remove any energy level
from the spectrum, but, instead, energy levels 
can, but not obligatorily,  be  introduced in its lower part.
Particular in special case  
of completely isospectral deformations of the (shifted) $L_m^{iso}$ systems,
all $m$ seed states composing negative scheme  
are  non-physical odd eigenstates of $L_0$.   
Analogous picture with dual negative schemes also is valid for rational
deformations of the harmonic oscillator system 
(\ref{DQHO}), where,  however,  
completely isospectral deformations of the (shifted) operator $L_0$ 
are impossible.

\subsection{Basic intertwining operators}
 
According to \cite{CarPly2,CIP}, 
with each of the dual schemes it is necessary 
first to associate two  basic pairs of the intertwining  operators.
Here, we discuss  general properties of such  operators 
without taking care of the concrete nature of the system built by the DCKA transformation. 
On the way, however, some important  distinctions between  rational deformations of
the AFF  model and harmonic  oscillator  have to be taken into account, and for this reason, 
it is convenient to speak of  two classes of the systems.  
We distinguish them by introducing the class index $c$,
where $c = 1$ and $c=2$ will correspond to deformed harmonic oscillator 
and  deformed AFF conformal mechanics model, respectively.

Although in essence both Hamiltonians $L_{(+)}$ and $L_{(-)}$ are the same operator, 
the corresponding DCKA transformations are very different from each other. 
The positive scheme generates the pair of Hermitian  conjugate  intertwining operators 
to be differential operators of order $n_+$, which for the sake of simplicity 
we denote as
$\A_{n_+}\equiv  A_{(+)}^-$ and $(A_{(+)}^-)^\dagger \equiv  A_{(+)}^+$.
In the same way, the negative scheme produces a different pair of intertwining operators, 
this time of order $n_-$,  
 $\A_{n_-}\equiv  A_{(-)}^-$ and $(A_{(-)}^-)^\dagger \equiv  A_{(-)}^+$.
 These operators satisfy intertwining relations
\begin{equation}
\label{inter0}
A_{(+)}^-L=L_{(+)}A_{(+)}^-\,, \qquad A_{(-)}^-L=L_{(-)}A_{(-)}^-\,,
\end{equation} 
and  
Hermitian conjugate  relations 
with 
$A_{(+)}^+$ and $A_{(-)}^+$,
where  $L$ is $L_0$ or $L_{0^+}$ in dependence on the class of 
the rationally deformed system $L_{(\pm)}$. 
In this way, operators 
$A_{(-)}^-$ and  $A_{(+)}^-$
transform eigenstates 
 of the harmonic oscillator 
into eigenstates  of $L_{(\pm)}$, but they 
do this in different ways. 
Applying operator identities 
  (\ref{inter0}) to an arbitrary physical or non-physical (formal)  eigenstate
 $\varphi_n$ of $L$ different from any seed state of the positive scheme 
 and using Eq. (\ref{L+L-}), one  can derive the equality
\be 
\label{relation-operators}
  A_{(-)}^-\varphi_n=A_{(+)}^-\varphi_{n+N}
\ee
to be valid modulo a multiplicative constant.
As a result, 
both operators acting on the same state of the harmonic oscillator produce 
different states
of the new system.
Analogous difference between  intertwining operators was 
already observed  in Section \ref{isoharconf} in the case of the 
non-deformed harmonic oscillator and 
AFF model.
The Hermitian conjugate operators  $A_{(-)}^+$ and  $A_{(+)}^+$ do a similar job but  
in the opposite direction. 
Eq.  (\ref{relation-operators}) suggests that some 
peculiarities 
 should be taken into account  for class 2 systems\,: 
 the infinite potential barrier  at $x=0$
 assumes that  
 physical states of $L_{0^+}$ and 
 $L_{(\pm)}$ systems
 are described by odd wave functions.
Then,
in order  $ A_{(+)}^-$ would transform 
physical states of $L_{0^+}$ into  physical states 
of $L_{(\pm)}$,  index $N$ has to be taken even in (4.6) for odd values of index $n$.
This means that 
$A_{(-)}^-$ transforms physical states into physical 
 only if $N$ is even.
In the case of odd $N$, it is necessary to take 
$A_{(-)}^-a^-$ or $A_{(-)}^-a^+$ as a physical  intertwining operator. 
It is convenient to take into account this peculiarity
by denoting 
the remainder of the division $N/c$ by 
$r(N,c)$\,: it takes value $1$ 
in the class $c=2$ of the systems with odd $N$ and equals zero
in all other cases. 

The alternate products of the described 
 intertwining operators are of the form (\ref{An+An-}), 
 and  
 for further analysis it 
is useful to write down  them explicitly: 
\begin{eqnarray}
\label{A-A-A+A+Poly}
&A_{(\pm)}^{+}A_{(\pm)}^{-}=P_{n_\pm}(L)\,,\qquad A_{(\pm)}^{-}A_{(\pm)}^{+}=P_{n_\pm}(L_{(\pm)})\,,&\\
\label{polyA}
&P_{n_+}(\eta)\equiv \prod_{k=1}^{n_+}(\eta-2l_k^+-1)\,, \qquad 
P_{n_-}(\eta)\equiv \prod_{k=1}^{n_-}(\eta+2l_k^-+1)\,.&
\end{eqnarray} 
Here $l_k^+$ are indexes of physical states of eigenvalues $2l_k^++1$
 in 
the set $\Delta_+$,
and $-l^-_k$ correspond to non-physical states of eigenvalues $-2l^-_k-1$ 
in 
the negative scheme $\Delta_-$.
 In the same 
 vein, it is useful to write 
$(a^+)^k(a^-)^k=T_k(L_0)$ and $(a^-)^k(a^+)^k=T_k(L_0+2k)$, where
\begin{eqnarray}
\label{Tk}
&T_{k}(\eta)\equiv \prod_{s=1}^{k}(\eta-2s+1)\,, \qquad T_{k}(\eta+2k)\equiv \prod_{s=1}^{k}(\eta+2s-1)\,.&
\end{eqnarray}
We also have the operator identities 
  \begin{eqnarray}
\label{ide}
(a^-)^{N}=(-1)^{n_-}A_{(-)}^+A_{(+)}^-\,,
\qquad f(L_{(-)})A_{(+)}^-(a^+)^{n_-}=(-1)^{n_-}h(L_{(-)})A_{(-)}^-(a^-)^{n_+}
\end{eqnarray}
and their Hermitian conjugate versions, 
where $f(\eta)$ and $h(\eta)$ are polynomials whose explicit structure 
is given in Appendix \ref{show}. 
In one-gap deformations of the harmonic oscillator and gapless deformations 
of $L_1^{iso}$ these polynomials reduce to $1$.

\subsection{Extended sets of ladder and intertwining operators}
\label{interladder}
The intertwining operators considered in the previous subsection 
allow us to construct ladder operators for the deformed systems, see
 \cite{CarPly2,CIP}\,: 
 \begin{itemize}
 \item [1.] Operators of the type $\mathfrak{A}$\,:  $\mathfrak{A}_{c}^\pm=A_{(-)}^-(a^\pm)^c A_{(-)}^+$,
 which have a structure of the  Darboux-dressed ladder operators of the harmonic oscillator $L_0$ ($c=1$)
 and of the half-harmonic oscillator $L_{0^+}$ ($c=2$).
 These operators detect all the separated states by annihilating them but act
 as ladder operators in the equidistant part of the spectrum.
 For gapless (completely isospectral) deformations of $L_{0^+}$ these are  
 the spectrum generating operators 
by which one can generate any physical eigenstate of the system from its any other physical eigenstate.

 \item[2.] Operators of the  $\mathfrak{B}$ type\,: $\mathfrak{B}_{c}^\pm=A_{(+)}^-(a^\pm)^c A_{(+)}^+$. 
 Their function is to identify each valence band\,:  the operators $\mathfrak{B}_{c}^+$  and $ \mathfrak{B}_{c}^-$
 annihilate, respectively, the highest and the lowest states in each valence band.
 In the equidistant part of the spectrum they also act as ordinary ladder operators. 
 For gapless deformations, this pair 
 is also a spectrum generating set, but  
 $|\mathfrak{B}_{2}^\pm| > |\mathfrak{A}_{2}^\pm|$,
 where $|.|$ denotes differential order of  operator.

\item [3.] Operators of the  $\mathfrak{C}$ type connect the equidistant part
of the spectrum with separated states. 
They are given by $\mathfrak{C}_{N+r(N,c)}^-=A_{(+)}^-(a^-)^{r(N,c)}A_{(-)}^+$
and 
$\mathfrak{C}_{N+r(N,c)}^+=A_{(-)}^-(a^+)^{r(N,c)}A_{(+)}^+$.
Index $N+r(N,c)$ in the notation for these ladder operators 
indicate that  their action is similar 
to the action of the operators  $(a^\pm)^{N+r(N,c)}$ 
in the case of the systems $L_0$ and $L_{0^+}$, respectively.
 \end{itemize} 
 
 As it was shown in 
\cite{CarPly2,CIP}, 
the minimal  complete set of the spectrum generating ladder operators
for deformed systems of both classes $c=1,2$ 
is formed by any of the two sets 
 $(\mathfrak{A}, \mathfrak{C})$  or  $(\mathfrak{B}, \mathfrak{C})$. 
 However, in  nonlinear algebras produced  by operators
 from any of these two sets, new structures are generated
of the nature similar to powers of $a^\pm$, see Appendix~\ref{apen-red}. 
 Because of this, 
 instead of talking about three types of the ladder operators, 
 it is convenient to talk about three families of the operators given by  
\begin{eqnarray}
\label{genlad}
&\mathfrak{A}_{k}^\pm\equiv A_{(-)}^-(a^\pm)^k A_{(-)}^+\,,\qquad \mathfrak{B}_{ k}^\pm\equiv A_{(+)}^-(a^\pm)^k A_{(+)}^+\,,
&\\
&\mathfrak{C}_{N\pm k'}^-\equiv A_{(+)}^-(a^\mp)^{k'}A_{(-)}^+ \,,\qquad
\mathfrak{C}_{N\pm k'}^+\equiv (\mathfrak{C}_{N\pm k'}^-)^\dagger\,,&\label{genlad+}
\end{eqnarray}
where, formally, $k$ can take any nonnegative integer value  and $k'$  is such that 
$N-k'\geq 0$, otherwise operators (\ref{genlad+})  reduce to $\mathfrak{A}_k^\pm$, 
see (\ref{C-}). At $k=0$ and $N-k'=0$ all these operators reduce to certain polynomials in 
 $L_{(\pm)}$.
Independently on the class of the system, or on whether 
the operators are physical or not, 
the three families  $\mathfrak{D}_{\rho,j}^\pm=( \mathfrak{A}_{j}^\pm, \mathfrak{B}_{j}^\pm,\mathfrak{C}_{j}^\pm )$,
$\rho=1,2,3$,  $j=1,2,\ldots$, satisfy the commutation relations of the form 
\begin{equation}
\label{sl2rh}
[L_{(\pm)},\mathfrak{D}_{\rho,j}^\pm]=\pm2j \mathfrak{D}_{\rho,j}^\pm \,,
\qquad [\mathfrak{D}_{\rho,j}^-,\mathfrak{D}_{\rho,j}^+]=\mathcal{P}_{\rho,j}(L_{(-)})\,, 
\end{equation}
where  $\mathcal{P}_{\rho,j}(L_{(-)})$ 
is a certain polynomial of the corresponding Hamiltonian operator of the system,
whose order  
as a polynomial   is equal to differential order 
of $\mathfrak{D}_{\rho,j}^\pm$ minus one,
see
Appendix  \ref{apen-red}.
Algebra
(\ref{sl2rh}) can be considered as a deformation of
 $\mathfrak{sl}(2,\R)$
 \cite{JM2}.
 \vskip0.1cm

Equation (\ref{sl2rh})  suggests that for physical operators 
the factor $2j$ has to be a multiple of the distance 
between two consecutive energy levels in the equidistant part of each system, 
that is $\Delta E=2c$.  Then, for $\mathfrak{A}$ and $\mathfrak{B}$ families, 
the physical operators are those 
whose  index is $j=lc$ with $l \in \N$, while
for $\mathfrak{C}$ family  index should be $j=N+r(N,c)+cs$, where $s$ is 
integer such that $j>0$. 

For class $c=2$ systems, 
the squares of non-physical operators of the $\mathfrak{C}$ type are physical and 
reduce 
to products  of other symmetry generators; see
Appendix \ref{apen-red}.
The unique special case is 
$\mathfrak{C}_{2N}^\pm=(-1)^{n_-}(\mathfrak{C}_{N}^\pm)^2$
with odd $N$, in which operators
$\mathfrak{C}_{2N}^\pm$ are  not secondary and have to be taken into account.

On the other hand, each ladder operator belonging to the
families $\mathfrak{A}$, $\mathfrak{B}$ and $\mathfrak{C}$ can be constructed 
by ``gluing" two  complementary generators of 
alternative   Darboux transformations.
Such generators
have
the form
$A_{(\pm)}^- (a^{\pm})^n$ with $n=cs$ and 
$A_{(\pm)}^- (a^{\mp})^n$ with $n=cs+r(N,c)$, where $s=0,1,2\ldots$, 
and  their physical nature in the sense 
of the comments related to  (\ref{relation-operators}) is
guaranteed by the indicated selection of values for parameter $n$; 
for $s=0$ we recover the basic 
intertwining operators.
{}From the point of view of DCKA transformations, 
factors  $(a^-)^n$  can be produced by selecting the set of eigenstates  $(0,1,\ldots,n-1)$
as  the seed states,
whose effect is to shift harmonic oscillator Hamiltonian operator  for the constant  
 $2n$. Analogously, factor $(a^+)^n$ corresponds to a negative scheme
 $(-0,-1,\ldots,-(n-1))$ which produces a shift for $-2n$.
 Despite these families of ladder and intertwining operators seem to be infinite,
one can use Eq.  (\ref{ide}) to reduce them 
to  finite sets   
of  operators.
The operators reducible to the products 
of other physical operators of  lower  order will not be considered 
by us here as basic elements of the set of generators of symmetry. 
In other words,  we admit  that some generators can appear 
 as coefficients in corresponding (super)algebraic relations.
The related  details are presented in Appendix 
\ref{apen-red}, and we describe below
the resulting picture.

First,  we turn to the sequences of operators 
(\ref{genlad}) and (\ref{genlad+}). 
For  gapless deformations of the AFF model, the 
spectrum generating 
set is given by any pair of the conjugate operators $\mathfrak{A}^{\pm}_{2}$, 
$\mathfrak{B}^{\pm}_{2}$, 
or $\mathfrak{C}^{\pm}_{2}$. In   
general case, only the operators
\begin{eqnarray}
\label{ladgen}
\left\{
\begin{array}{cc}
\mathfrak{A}_{k}^\pm\,, & 0<k<N\,,\\
\mathfrak{B}_{k}^\pm\,, & 0<k<N\,,\\
\mathfrak{C}_{k}^\pm\,, & 0<k<2N+r(N,c)\,,\\
\end{array}
\right.
\end{eqnarray}
are the basic symmetry generators while the other 
can be written in terms of them  and polynomials in $L_{(\pm)}$.    
For  one-gap deformations of the  harmonic 
oscillator, the set of basic ladder operators can be reduced further
to the set
\begin{eqnarray}
\label{basicsubsetonegap}
\left\{
\begin{array}{cc}
\mathfrak{A}_{k}^\pm\,, & 0<k<n_+\,,\\
\mathfrak{B}_{k}^\pm\,, & 0<k<n_-\,,\\
\mathfrak{C}_{k}^\pm\,, & M<k<n_+\,,\\
\end{array}
\right. \qquad 
M=\left\{\begin{array}{ccc}
\max\,(n_-,n_+) & \text{if} & n_-\neq n_+\,,\\
N/2 & \text{if} & n_-=n_+\,,
\end{array}
\right.
\end{eqnarray}
and also we have relations 
$\mathfrak{A}_{n_+}^{\pm}=(-1)^{n_-}\mathfrak{C}_{n_+}^{\pm}$ and  $\mathfrak{B}_{n_-}^{\pm}=(-1)^{n_-}\mathfrak{C}_{n_-}^{\pm}$.

Consider now the issue of reduction of sequences of the intertwining operators.
For general deformations,  
only  the 
operators
\begin{eqnarray}
\label{genA}
\left\{
\begin{array}{cc}
A_{(\pm)}^-(a^\pm)^{n}\,, & 0\leq n<N\,,\\
A_{(\pm)}^-(a^\mp)^{n}\,,& 0<n<N+r(N,c)\,,
\end{array}
\right.
\end{eqnarray}
and their Hermitian conjugate counterparts 
can be considered as basic, see Appendix \ref{apen-red}.
One can note 
that the total number of the basic intertwining operators $\#_f=2[(4N-2+r(N,c))/c]$ 
is greater than the number of the basic ladder operators 
$\#_{lad} = 2[(4N-3+r(N,c))/c]$ 
which can be constructed  with their help. 
 In particular case of gapless deformations of the AFF model, 
 the indicated set of Darboux generators  can be reduced to 
 those which produce, by  `gluing' procedure, one conjugate pair of the 
 spectrum generating 
 ladder operators of the form
$\mathfrak{D}_{2,\rho}^\pm$.

For $c=1$ one-gap systems, identity
(\ref{ide}) allows us to reduce further the set of the basic intertwining 
operators, which, together with corresponding Hermitian conjugate operators, 
 is  given by any of the two options,  
\begin{equation}
\label{frakS}
\mathfrak{S}_{z}\equiv \left\{ \begin{array}{lcc}
              A_{(-)}^-{(a^+)}^{|z|}\,,  &  -N<z\leq 0\,,\\\vspace{-0.4cm}
             \\ A_{(-)}^-{(a^-)}^{z}\,,  &  0< z \leq n_+\,,\\\vspace{-0.4cm}
             \\ A_{(+)}^-{(a^+)}^{N-z} \,,  & n_+ < z \leq N\,,\\\vspace{-0.4cm}
             \\ A_{(+)}^-{(a^-)}^{N-z}\,,   & N<z<2N\,, \\
             \end{array}
   \right.\\\qquad \text{or}\qquad \mathfrak{S}_{z}^{'}\equiv \mathfrak{S}_{N-z}\,,
\end{equation}
 see Appendix \ref{apen-red}. 
Here we have reserved  $z=0$ and $z=N$ values for index $z$ to the dual schemes'
intertwining operators: in the first choice, $\mathfrak{S}_{0}=A_{(-)}^-$ and
$\mathfrak{S}_{N}=A_{(+)}^-$, and  
for the second choice we have $\mathfrak{S}_{0}'=A_{(+)}^-$ 
and $\mathfrak{S}_{N}'=A_{(-)}^-$. Written in this way, these operators satisfy the  
intertwining relations $
 \mathfrak{S}_{z}L=(L_{(-)}+2z)\mathfrak{S}_{z}$ or $\mathfrak{S}_{z}'L=(L_{(+)}-2z)\mathfrak{S}_{z}'$,
and their Hermitian conjugate versions. 
Then,  to study supersymmetry, we have to choose   either positive or negative scheme
to define the  $\mathcal{N}=2$ superextended Hamiltonian.
We take $\mathfrak{S}_{z}$ if we work with a negative scheme, 
and $\mathfrak{S}_{z}'$ if positive scheme is chosen for the construction of  super-extension.

\subsection{Supersymmetric extension}
\label{susyextension}
For each of the two dual schemes, one can construct an $\mathcal{N}=2$ superextended 
Hamiltonian operator of the form (\ref{Hlambda*}) by choosing appropriately $H_1=\breve{L}-\lambda_*$
and $H_0=L-\lambda_*$.
We put $\breve{L}=L_{(+)}$ and $\lambda_*=\lambda_+= 2l_1^++1$
for positive scheme, and choose  $\breve{L}=L_{(-)}$ and $\lambda_*=\lambda_-= -2l_1^--1$
for negative scheme.
For  both options, we set $L=L_0$ if we are dealing with a rational extension of 
harmonic oscillator, and $L=L_{0^+}$ if we  work with a deformation of the AFF model.
 We name the matrix Hamiltonian associated with  negative scheme as $\mathcal{H}$, and
denote by $\mathcal{H}'$ the Hamiltonian  of positive scheme. 
The  spectrum of  these systems can be  found using 
 the properties of  the corresponding  intertwining operators described in
 Section \ref{Dar}, see also refs.  \cite{CarPly2,CIP}.
The two  Hamiltonians are connected by relation 
 $\mathcal{H}-\mathcal{H}'=-N(1+\sigma_3)-\lambda_-+\lambda_+$,
 and  $\sigma_3$ plays a role of 
 the $\mathcal{R}$ symmetry
generator  for both  superextended systems.
In this subsection we finally construct the corresponding 
spectrum generating superalgebra for $\mathcal{H}$  and  $\mathcal{H'}$. 
The resulting structures are based on the physical operators $\mathfrak{D}_{\rho,j}^\pm$.
As we shall see, 
the supersymmetric versions of the  $c=1$ systems
are described by a nonlinearly extended super-Schr\"odinger symmetry with 
bosonic  generators  to be differential operators of even and odd orders, while in the case 
of the $c=2$ systems we obtain
nonlinearly  extended superconformal symmetry in which bosonic generators 
are of even order only.  
\vskip0.1cm

We construct a pair of fermionic operators on the basis 
of each intertwining operator from the set (\ref{genA}) and 
its Hermitian conjugate counterpart.  
Let us consider first 
the extended nonlinear super-Schr\"odinger symmetry of a one-gap deformed 
harmonic oscillator,
and then we generalize the picture.  
If we choose the negative scheme, then we use $\mathfrak{S}_z$ defined in (\ref{frakS}) to construct  
the set of operators 
\begin{eqnarray}
\label{gencharge}
\mathcal{Q}_1^{z}=
\left(
\begin{array}{cc}
  0&  \mathfrak{S}_z  \\
 \mathfrak{S}_z^\dagger &  0     
\end{array}
\right)\,,
\qquad 
\mathcal{Q}_2^{z}= i\sigma_3\mathcal{Q}_1^{z}\,, \qquad -N<z<2N\,.
\end{eqnarray}
They satisfy  the (anti)-commutation relations 
\begin{equation}
\label{SUSY}
[\mathcal{H},\mathcal{Q}_a^z]=2iz\epsilon_{ab}\mathcal{Q}_{b}^{z}\,,
\qquad \{\mathcal{Q}_a^z,\mathcal{Q}_b^z\}=2\delta_{ab}\mathbb{P}_{z}(\mathcal{H},\sigma_3)\,,\qquad
[\Sigma,\mathcal{Q}_{a}^z]=-i\epsilon_{ab}\mathcal{Q}_{b}^{z}\,,
\end{equation}
where $\mathbb{P}_z$ are some polynomials whose structure is described in 
Appendix \ref{apen-comm}. 
For the choice of the positive scheme to fix extended Hamiltonian, according to 
(\ref{frakS}), the  corresponding fermionic operators are given by 
$\mathcal{Q}_1^{'z}\equiv \mathcal{Q}_1^{N-z}$. They satisfy relations 
of the same form (\ref{SUSY}) but with replacement 
$\mathcal{H}\rightarrow \mathcal{H}'$, $\Sigma=\frac{1}{2}\sigma_3 \rightarrow \Sigma'=-\frac{1}{2}\sigma_3$, 
$\mathbb{P}_{z}(\mathcal{H},\sigma_3)\rightarrow 
\mathbb{P}_{z}'(\mathcal{H}',\sigma_3)=\mathbb{P}_{N-z}(\mathcal{H}'-N(1+\sigma_3)-\lambda_-+\lambda_+,\sigma_3)$,
$\mathcal{Q}_{1}^{z}\rightarrow \mathcal{Q}_{2}^{'z}$ and
$\mathcal{Q}_{2}^{z}\rightarrow \mathcal{Q}_{1}^{'z}$. 
The  fermionic operators $\mathcal{Q}_a^{0}$ (or $\mathcal{Q}_a^{'0}$) 
are the  supercharges 
of the (nonlinear in general case) $\mathcal{N}=2$ Poincar\'e supersymmetry,
which are 
integrals of motion of the system $\mathcal{H}$ 
(or $\mathcal{H}'$), and 
 $\mathbb{P}_{0}=
P_{n_-}(\mathcal{H} + \lambda_{-})\,$
 (or $\mathbb{P}_{0}=P_{n_+}(\mathcal{H}' +
\lambda_{+})$)  with polynomials   $P_{n_\pm}$ defined in (\ref{polyA}).
The operators $\mathcal{Q}_a^{'0}$ are analogous here to supercharges 
in (\ref{supercharge1}).
On the other hand, we have here the  
fermionic operators $\mathcal{Q}_a^{'N}$ as  analogs 
of dynamical integrals $\mathcal{S}^a_\nu$ there. 
We recall that in the  simple linear case considered in Section  \ref{isoharconf}, 
the interchange between positive and negative  schemes
corresponds to the  automorphism of superconformal algebra,
and this observation will be helpful for us  
for the analysis of the nonlinearly extended
super-Schr\"odinger
structures.
Here, actually, 
each of the  $(\#_f-2)/2$ pairs of fermionic operators distinct from supercharges
provides a possible dynamical extension of the super-Poincar\'e symmetry.
As we will see,  all of them are necessary to obtain a closed
nonlinear 
spectrum generating 
superalgebra of the superextended system.

To construct any extension of the deformed Poincar\'e supersymmetry, 
we calculate $\{\mathcal{Q}_{a}^{0},\mathcal{Q}_{a}^{z} \}$,  in the negative scheme, 
or $\{\mathcal{Q}_{a}^{'0},\mathcal{Q}_{a}^{'z} \}$ in the positive one.
In the first case we have
\be 
\label{Cn+kQN}
\{\mathcal{Q}_a^{0},\mathcal{Q}_{b}^{z}\}=\delta_{ab}(\mathcal{G}_{-z}^{(2\theta(z)-1)}+ 
\mathcal{G}_{+z}^{(2\theta(z)-1)})+i\epsilon_{ab}(\mathcal{G}_{-z}^{(2\theta(z)-1)}- \mathcal{G}_{+z}^{(2\theta(z)-1)})\,,
\ee
where $z\in (-N,0)\cup(0,2N)$, $\theta(z)=1\, (0)$ for $z>0\, (z<0)$, 
 and  $\mathcal{G}^{(2\theta(z)-1)}_{\pm z}$
are given by 
\be
\label{superC}
\mathcal{G}_{+z}^{(2\theta(z)-1)}=
\left(
\begin{array}{cc}
 \mathfrak{S}_{0}(\mathfrak{S}_{z})^{\dagger} &  0  \\
 0&     (\mathfrak{S}_{z})^{\dagger}\mathfrak{S}_{0}   
\end{array}
\right)\,,\qquad
\mathcal{G}_{-z}^{(2\theta(z)-1)}=(\mathcal{G}_{+z}^{(2\theta(z)-1)})^\dagger\,.
\qquad 
\ee 
Following  definition (\ref{frakS}), one finds  directly that 
 $\mathfrak{S}_{0}(\mathfrak{S}_{z})^{\dagger}$ is equal to $\mathfrak{A}_{|z|}^-$ when $-N<z<0$, 
 while  for $0<z\leq n_+$, this operator is equal to $\mathfrak{A}_{z}^+$,  
and takes the form of  $\mathfrak{C}_{z}^+$ for $n_+<z<2N$.
The operators $(\mathfrak{S}_{z})^{\dagger}\mathfrak{S}_{0}$
reduce to
\be
\label{S*S+}
(\mathfrak{S}_{z})^{\dagger}\mathfrak{S}_{0}= \left\{ \begin{array}{lcc}
              P_{n_-}(L-2k)(a^-)^{|z|}\,, & -N<z<0\,,\\\vspace{-0.4cm}
             \\(a^+)^{z}P_{n_-}(L)\,,   &  0<z\leq n_+\,,\\\vspace{-0.4cm}
             \\ (-1)^{n_-}(a^+)^zT_{N-z}(L+2N)\,,  &  n_+<z<N\,,\\\vspace{-0.4cm}
             \\ (-1)^{n_-}(a^+)^z \,,  & N\leq z<2N\,.
             \end{array}
   \right.\\
\ee
Note that $\mathcal{G}_{\pm k}^{(-1)}$ and $\mathcal{G}_{\pm k}^{(+1)}$ with $k=|z|\leq n_-$ 
are two different matrix extensions of  the same operator $\mathfrak{A}_k^\pm$.

For a superextended system based on  the positive scheme, we obtain     
\be 
\label{Q0'Qk'}
\{\mathcal{Q}_a^{'0},\mathcal{Q}_{b}^{'z}\}=
\delta_{ab}(\mathcal{G}^{'(2\theta(z)-1)}_{-z}+\mathcal{G}^{'(2\theta(z)-1)}_{+z})
-i\epsilon_{ab}(\mathcal{G}^{'(2\theta(z)-1)}_{-z}- \mathcal{G}^{'(2\theta(z)-1)}_{+z})\,,
\ee
where, again,  $z\in (-N,0)\cup(0,2N)$, and  $\mathcal{G}^{'(2\theta(z)-1)}_{\pm z}$
are given by  
\be
\label{superC'}
\mathcal{G}^{'(2\theta(z)-1)}_{-z}=
\left(
\begin{array}{cc}
 \mathfrak{S'}_{0}(\mathfrak{S'}_{z})^{\dagger} &  0  \\
 0&     (\mathfrak{S'}_{z})^{\dagger}\mathfrak{S'}_{0}   
\end{array}
\right)\,,\qquad
\mathcal{G}^{'(2\theta(z)-1)}_{+z}=(\mathcal{G}^{'(2\theta(z)-1)}_{-z})^\dagger\,.
\qquad 
\ee 
Now,  
$\mathfrak{S'}_{0}(\mathfrak{S'}_{z})^{\dagger}=\mathfrak{B}^{+}_{|z|}$ when $-N<z<0$,
while for positive index $z$ this operator reduces to  
$\mathfrak{B}^{-}_{z}$ 
 when $0<z\leq n_-,$ and to $\mathfrak{C}^{-}_{z}$
 when $n_-<z<2N$. For the other matrix element we have
 \begin{eqnarray}
(\mathfrak{S}_{z}')^{\dagger}\mathfrak{S}_{0}'= \left\{ \begin{array}{lcc}
(a^+)^{|z|}P_{n_+}(L)\,, & -N<z<0\,,\\\vspace{-0.4cm}
              \\(a^-)^{z}P_{n_+}(L)\,, &  0<z\leq n_-\,,\\\vspace{-0.4cm}
             \\ (-1)^{n_-}T_{N-k}(L)(a^-)^z\,, & n_-<z<N\,,\\\vspace{-0.4cm}
             \\ (-1)^{n_-}(a^-)^z\,, & N<z<2N\,.
             \end{array}
   \right. 
 \end{eqnarray} 
Here, again, there are two different matrix extensions of the 
operators of the $\mathfrak{B}$-family given by 
$\mathcal{G}_{\pm k}^{'(+1)}$ and $\mathcal{G}_{\pm k}^{'(-1)}$ when $k\leq n_-$. 

By comparing both schemes 
one can note two other special features.
It turns out  that $\mathcal{G}_{\pm k}^{(1)}=\mathcal{G}_{\pm k}^{'(1)}$ when 
$k\geq N$, and this corresponds to  the automorphism discussed 
in Section~\ref{isoharconf}. In the same way,
for $\max(n_-,n_+)<k<N$, operators $\mathcal{G}_{\pm k}^{(1)}$ and $\mathcal{G}^{'(1)}_{\pm k}$
are different matrix extensions of  $\mathfrak{C}^{\pm}_k$. 

{}From here and in what follows 
we do not specify whether we have the superextended system 
corresponding to the negative or the positive scheme, and will just use, respectively, 
 the unprimed or primed 
notations for operators   of the alternative dual schemes. 
In particular, we have
\begin{equation}
[\mathcal{H},\mathcal{G}_{\pm k}^{(2\theta(z)-1)}]=\pm 2k \mathcal{G}_{\pm k}^{(2\theta(z)-1)}
\,,\qquad k\equiv|z|\,, \qquad z \in (-N,0)\cup(0,2N)\,,
\end{equation} 
that shows explicitly that our new bosonic operators have the nature of  ladder operators 
of the superextended system $\mathcal{H}$.  
Commutators $[\mathcal{G}_{-k}^{(1)},\mathcal{G}_{+k}^{(1)}]$ and 
$[\mathcal{G}_{-k}^{(-1)},\mathcal{G}_{+k}^{(-1)}]$ 
produce polynomials in $\mathcal{H}$ and $\sigma_3$,
 which can be calculated by using the polynomials  
$\mathcal{P}_{\rho,j}$ defined in  (\ref{sl2rh}). 
The algebra generated by  $\mathcal{H}$, $\mathcal{G}_{\pm k}^{(2\theta(z)-1)}$
and $\sigma_3$ is  identified 
as  a deformation of $\mathfrak{sl}(2,\R)\oplus \mathfrak{u}(1)$,  
where a concrete form of deformation depends on 
the system, $\mathcal{H}$,  and  on $z$.
Each of these nonlinear 
bosonic algebras expands further  up to  
a certain closed nonlinear deformation of superconformal  $\mathfrak{osp}(2|2)$ algebra
generated by the  subset
of operators
\begin{equation}
\label{U1}
\mathcal{U}_{0,z}^{(2\theta(z)-1)}\equiv \{\mathcal{H},\sigma_3,
\mathbb{I},\mathcal{G}_{\pm |z|}^{(2\theta(z)-1)}, \mathcal{Q}_{a}^{0},
\mathcal{Q}_{a}^{z}\}\,,\qquad z \in (-N,0)\cup(0,2N)\,,
\end{equation} 
see Appendix \ref{apen-comm}.

The deficiency of  any of these nonlinear superalgebras 
is that none of them is a spectrum generating algebra for the
superextended system\,: application 
of operators from the set (\ref{U1}) and of their products 
does not allow to connect two arbitrary eigenstates in the spectrum 
of $\mathcal{H}$. 
To find the spectrum generating superalgebra  
for this kind of the superextended systems, one can try 
 to include into the superalgebra simultaneously 
 the operators  $\mathcal{G}^{(1)}_{\pm N}$
 and, say, $\mathcal{G}^{(1)}_{\pm 1}$ or $\mathcal{G}^{(-1)}_{\pm 1}$.
 The operators  $\mathcal{G}^{(1)}_{\pm N}$ 
 provide us with matrix extension of the operators $\mathfrak{C}_{N}^\pm$
 being ladder operators for deformed subsystems $L_{(-)}$ or  $L_{(+)}$.
 Analogously, operators $\mathcal{G}^{(1)}_{\pm 1}$ or $\mathcal{G}^{(-1)}_{\pm 1}$
 supply us with matrix extensions  of the ladder operators 
$\mathfrak{A}_{ 1}^\pm$ or $\mathfrak{B}_{ 1}^\pm$ 
($\mathfrak{A}_{ 2}^\pm$ or $\mathfrak{B}_{ 2}^\pm$)
when systems $L_{(\pm)}$ are of the class $c=1$ or $c=2$ with even (odd) $N$.
Therefore, it is enough to unify  the sets of generators 
  $\mathcal{U}_{0,1}^{(1)}$ and  $\mathcal{U}_{0,N}^{(1)}$. 
 Having in mind the commutation relations 
 between operators of the three families  
$\mathfrak{A}$, $\mathfrak{B}$ and $\mathfrak{C}$, one can find, however, 
that the commutators of  the operators 
 $\mathcal{G}^{(1)}_{\pm N}$  with $\mathcal{G}^{(1)}_{\pm 1}$
 generate other bosonic matrix operators  $\mathcal{G}^{(1)}_{\pm k}$.
 The commutation of these operators with supercharges  $\mathcal{Q}_{a}^{0}$
 generates the rest of the fermionic operators we considered, see
 Appendix \ref{apen-comm}  for details.
The set of higher order generators 
is completed by considering all non-reducible bosonic and fermionic generators,
which do not decompose into the products of other generators.
In correspondence with that was noted above, we 
 arrive finally at two different extensions of the sets of 
operators with index less than $N$. By this reason it is convenient also 
to  introduce   the operators 
\begin{eqnarray}
\label{G_k}
&\mathcal{G}_{\pm k}^{(0)}\equiv \Pi_- (a^\pm)^k, \qquad
 k=1,\ldots,N-1, \qquad 
 \Pi_-=\frac{1}{2}(1-\sigma_3)\,,&
\end{eqnarray} 
which help us to fix in a unique way the bosonic set of generators. For our purposes we choose 
to write all the operators $\mathcal{G}_{\pm k}^{(-1)}$ in terms of $\mathcal{G}_{\pm k}^{(1)}$ and 
$\mathcal{G}_{\pm k}^{(0)}$ when $k\leq n_+$ in the negative scheme, 
and when $k\leq n_-$ in the extended system associated with the positive scheme.
For indexes outside the indicated scheme-dependent range, we neglect operators $\mathcal{G}_{\pm k}^{(-1)}$ 
 because they are not basic in correspondence with  discussion on reduction of ladder operators in the 
 previous  subsection \ref{interladder}. 
As a result, we have to drop out in  (\ref{U1}) all the operators $\mathcal{G}_{\pm |z|}^{(2\theta(z)-1)}$ with 
$z\in (-N,0)$.

By taking  anti-commutators  of fermionic operators $\mathcal{Q}_{a}^{N}$ 
with $\mathcal{Q}_{a}^{z}$, $z\neq 0$, we produce bosonic
dynamical integrals $\mathcal{J}_{\pm |z-N|}^{(1-2\theta(z-N))}$,
which  have exactly the same 
structure of  the even generators 
$\mathcal{G}_{\pm |z|}'^{(2\theta(z)-1)}$
in the extension associated with  the dual scheme.
In this way we obtain the subsets of 
operators
\begin{eqnarray}
\label{I1}
\mathcal{I}_{N,z}^{(1-2\theta(z-N))}\equiv \{\mathcal{H},\sigma_3,\mathbb{I},
\mathcal{J}_{\pm |z-N|}^{(1-2\theta(z-N))}, \mathcal{Q}_{a}^{N},\mathcal{Q}_{a}^{z}\}\,
\qquad z \in (-N,0)\cup(0,2N)\,,
\end{eqnarray}
which also generate closed nonlinear super-algerabraic structures. 
With the help of  (\ref{G_k}), we find similarly to the subsets (\ref{U1}),
that a part of the sets  (\ref{I1}) also can be reduced. 

Having  in mind the ordering relation between $n_-$ and $n_+$, 
the superextended systems associated with 
the negative schemes 
can be   characterized finally by the following irreducible, 
in the sense of subection \ref{interladder},
subsets of symmetry generators\,:
\begin{table}[htbp]
\begin{center}
\begin{tabular}{ll}
\hline
$n_-\leq n_+$& $n_+<n_- $\\\hline 
$\mathcal{U}_{0,k}^{(1)}\,,  0<k<2N$ & $\mathcal{U}_{0,k}^{(1)}\,, k\in (0,n_+)\cup (n_-,2N)$ \\
$\mathcal{I}_{N,z}^{(1-2\theta(N-z))}\,,  z\in(-N,0)\qquad\qquad$ & $\mathcal{I}_{N,z}^{(1-2\theta(N-z))}\,,  z\in(-N,0)\cup$\\
$\cup(n_+,N)$                                             & $[n_+,N)$
 \\\hline
\end{tabular}
\end{center}
\end{table}
\noindent
For more details, 
see  Appendix \ref{apen-red}. 
A similar result can be obtained for superextended systems
associated with   positive schemes, where
the roles played by  families $\mathfrak{A}$ and $\mathfrak{B}$, 
and of numbers $n_-$ and $n_+$
are interchanged.

Finally, we arrive  at the following picture. Any operator that can be obtained via 
(anti)commutation relations of the basic generators from the above Table but does not belong to them   
can be written as their product. 
As a result, the listed basic operators produce the spectrum generating superalgebra
for  one-gap rationally deformed super-extended systems. 
It is worth it to stress that in this set of generators the unique 
true integrals of motion, in addition to $\mathcal{H}$ and $\sigma_3$,
 are the supercharges $\mathcal{Q}_a^{0}$, while the rest  has to
 be promoted to the dynamical integrals by using transformation
 (\ref{recipeintegrals}).

For gapless rational extensions of the systems of class $c=2$, 
only the subset $\mathcal{U}_{0,2}^{(1)}$ has to be considered
instead of the family of sets $\mathcal{U}_{0,k}^{(1)}$. 
For super-extensions of rationally deformed systems of arbitrary form in the sense of the class $c$ 
and arbitrary number of gaps and their dimensions,  
the identification of their generalized super-Schr\"odinger or superconformal 
structures  is realized in a similar way. The procedure is based
on  the sets of operators (\ref{ladgen}) and 
(\ref{genA}), which include  the operators 
 (\ref{basicsubsetonegap})  and  (\ref{frakS})
 of the discussed one-gap case as subsets.
 As a result, 
for every irreducible pair of ladder operators (\ref{ladgen}) with index less than $N$
we have two super-extensions which are related  by operators of the form (\ref{G_k}). When we 
put together the 
subsets containing the spectrum generating set of operators, 
we obtain all the other structures.

\section{Examples
}
\label{examples}
 
In this section we employ the developed machinery 
for two simple examples of rationally extended systems.
We consider here the  cases  in which  the negative scheme is characterized by only 
one seed state,  $n_-= 1$, 
 and  then  $N=n_++1$. 
The systems  generated by Darboux transformations of this kind could be a one-gap rational extension 
of the harmonic oscillator, or a gapless deformation of $L_{1}^{iso}$, and according 
to the general picture  described in the previous section, 
one can 
 make some 
general assertions. 
\vskip0.1cm
\begin{itemize}
\item[(i)]\emph{Peculiarities of one-gap deformations of the QHO}\,: 
The superextended Hamiltonian constructed on the
base of the negative scheme with $n_-=1$ 
is characterized by unbroken  
$\mathcal{N}=2$ 
Poincar\'e supersymmetry whose 
supercharges, being  the first 
order differential operators, generate
a  Lie superalgebra.  
The $\mathfrak{B}$ family of ladder operators in the sense of (\ref{basicsubsetonegap}) 
does not play any 
role in this scheme. 
On the other hand, the super-Hamiltonian provided by the positive scheme
possesses  $n_+$ 
singlet states while the ground state is a doublet. 
The $\mathcal{N}=2$ super-Poincar\'e 
algebra of such a system  is nonlinear
as its supercharges are of differential order 
$n_+=2\ell \geq 2$.  
\vskip0.1cm

\item[(ii)]\emph{Peculiarities of gapless deformations of $L_1^{iso}$}\,:
The negative scheme produces a super-Hamiltonian with spontaneously broken 
supersymmetry, whose all energy levels are doubly degenerate;
its $\mathcal{N}=2$  super-Poincar\'e algebra  has linear nature. 
To construct the spectrum generating algebra we only need a matrix extension
of the operators $\mathfrak{A}_2^\pm$. In a superextended system produced by
the positive scheme,  $n_+>1$ physical and non-physical states of $L_{0^+}$ of positive energy
(the latter being even eigenstates of harmonic oscillator)
are used as seed states for Darboux transformation.
Its supersymmetry is spontaneously broken, 
and the $\mathcal{N}=2$ super-Poincar\'e algebra is nonlinear. 
The nonlinearly deformed super-Poincar\'e symmetry cannot be expanded 
up to spectrum generating superalgebra by combining it with matrix 
extension  of  the $\mathfrak{A}^\pm_2$, 
but this can be done by using matrix extensions of the 
$\mathfrak{B}_2^\pm$ or $\mathfrak{C}_2^\pm$ ladder operators,
see  (\ref{superC'}). 
The resulting  spectrum generating superalgebra is a certain nonlinear 
deformation 
of the  $\mathfrak{osp}(2|2)$ superconformal symmetry.     
\vskip0.1cm

\end{itemize} 

In what follows, 
we  first  consider the simplest example of the deformed gapless AFF model, and 
then we discuss the super-extension  with a rational one-gap deformation  
of the harmonic oscillator. As we shall see, in the second case
the algebraic structure of the deformed superextended Schr\"odinger symmetry
is  more complicated than that of the deformed superconformal
symmetry associated with gapless deformation of the AFF system.

\subsection{Gapless deformation of AFF model}

The super-extension  we consider here is based on the 
dual schemes $(1,2,3)\sim (-3)$ composed  from 
eigenstates of the half-harmonic oscillator $L_{0^+}$. 
We have  here $N=4$,
and seed states $\psi_2$ and $\psi_{-3}$ are non-physical.
The Darboux transformation of  the negative scheme generates
 the Hamiltonian 
\begin{eqnarray}
\label{L(-3)}
&L_{(-)}\equiv  L_{(-3)}=-\frac{d^2}{dx^2}+x^2-2(\ln \psi_{-3})''=
L_{1}^{iso}+8 \frac{2 x^2-3}{(3 + 2 x^2)^2}-2\,,&
\end{eqnarray}
and the pair of the corresponding intertwining operators is
\begin{eqnarray}
\label{A-3}
&A_{(-)}^-\equiv  A_{(-3)}^-=A_{-1}^--\frac{4x}{2x^2+3}\,,\qquad 
A_{(-)}^+\equiv  A_{(-3)}^+=A_{-1}^+-\frac{4x}{2x^2+3}\,.&
\end{eqnarray}
For them the second intertwining  relation from  (\ref{inter0})  and its Hermitian conjugate version are valid. 
Following the way described in the previous section, we obtain
$P_{n_-}(L_{(-)})=L_{(-)}+7\equiv  H_1$  and   
$P_{n_-}(L_{0^+})=L_{0^+}+7\equiv  H_0$. 
Physical  eigenstates $\phi_l$ of the deformed  system $H_1$ are obtained 
from odd eigenfunctions
of $H_0$,  $\phi_l=A_{(-)}^-\psi_{2l+1}$, 
$l=0,1,2,\ldots$. The state $\phi_l=A_{(-)}^-\widetilde{\psi_{(-3)}}=1/\psi_{(-3)}$ 
diverges in  $x=0$, and therefore is not a physical state of  $L_{(-)}$. 
Since the state
$\psi_{-3}$ is not physical,  $H_1$ is completely  isospectral to $H_0$.
In fact, this is the simplest example of this category.

In positive scheme we have the intertwining operators  
$A_{(+)}^\pm=A_{(1,2,3)}^\pm$, which are constructed iteratively 
using three seed states according to the prescription (\ref{Andef}).
These operators satisfy the first intertwining relation from  
 (\ref{inter0}) as well as its conjugate version,
 and their products are  
 $P_{n_+}(L_{0^+})=(L_{0^+}-3)(L_{0^+}-5)(L_{0^+}-7)$
 and  
$P_{n_+}(L_{(+)})=(L_{(+)}-3)(L_{(+)}-5)(L_{(+)}-7)$,
where we use Eq.  
(\ref{A-A-A+A+Poly}) and $L_{(+)}=L_{(-)}+8$.

Consider now the superextended system constructed on the base 
of the negative scheme. 
At the end of this subsection we discuss  the super-extension 
based on the positive scheme.
The superextended Hamiltonian and its spectrum are given by
\begin{equation}
\label{hamilisodef}
\mathcal{H}=
\left(
\begin{array}{cc}
 H_1&    0 \\
0 &  H_0    
\end{array}\right),\qquad \mathcal{E}_{n}=4n+10\,,
\qquad n=0,1,\ldots\,.
\end{equation}
Due to complete isospectrality  of $ H_1$ and  $H_0$, all the energy levels
of the system (\ref{hamilisodef}) 
including the  lowest one $\mathcal{E}_{0}=10>0$ 
are  doubly  degenerate. The  Witten's index 
equals zero, and we have here the case 
of spontaneously broken $\mathcal{N}=2$ super-Poincar\'e symmetry 
generated by Hamiltonian $\mathcal{H}$, the supercharges $\mathcal{Q}^0_a$
constructed in terms of $A_{(-)}^\pm$,
and by  $\Sigma=\frac{1}{2}\sigma_3$.

The complete set of fermionic operators $\mathcal{Q}_a^{z}$ with $-3\leq z\leq 7$ is given by
Eq.  (\ref{gencharge}),  and the ladder operators  $\mathcal{G}_{\pm k}^{(1)}$ with $k=1,\ldots,7$
are constructed according to  (\ref{superC}). 
However,  
due to the isospectrality of the subsystems, we need here only the restricted set 
of operators
$\mathcal{U}_{0,2}^{(1)}=\{\mathcal{H},\mathbb{I},\mathcal{G}_{\pm2}^{(1)},\sigma_3,\mathcal{Q}_a^{0},\mathcal{Q}_a^{2} \}$ 
to obtain the spectrum generating superalgebra of the system,
where 
\begin{eqnarray}
\label{Q2Cpm2}
\mathcal{Q}^z_1=
\left(
\begin{array}{cc}
 0&    A^-_{(-)}(a^-)^z \\
(a^+)^zA^+_{(-)} &  0    
\end{array}\right),\,\, z=0,2\,;\quad
\mathcal{G}_{-2}^{(1)}=
\left(
\begin{array}{cc}
 A_{(-)}^-(a^-)^2A^+_{(-)}&   0 \\
0 &  H_0(a^-)^2   
\end{array}\right),
\end{eqnarray}
and $\mathcal{Q}^z_2=i\sigma_3 \mathcal{Q}^z_1$, $\mathcal{G}_{+2}^{(1)}=(\mathcal{G}_{-2}^{(1)})^\dagger$.
They satisfy the superalgebraic relations
\begin{eqnarray}
\label{nonlinear3}
&[\mathcal{H},\mathcal{Q}_a^{0}]=0\,,\qquad 
[\mathcal{H},\mathcal{Q}_a^{2}]=4i\epsilon_{ab}\mathcal{Q}_b^{2}\,,\qquad
[\sigma_3,\mathcal{Q}_a^{z}]=-2i\epsilon_{ab}\mathcal{Q}_b^{z}\,,\quad z=0,2\,,&\\
\label{nonlinear2}
&\{\mathcal{Q}_a^{0},\mathcal{Q}_a^{0}\}=2\delta_{ab}\mathcal{H}\,,\qquad
\{\mathcal{Q}_a^{0},\mathcal{Q}_b^{2}\}=\delta_{ab}(\mathcal{G}_{-2}^{(1)}+\mathcal{G}_{+2}^{(1)})+i
\epsilon_{ab}(\mathcal{G}_{-2}^{(1)}-\mathcal{G}_{+2}^{(1)})\,,&\\
\label{nonlinear1}
&[\mathcal{H},\mathcal{G}_{\pm 2}^{(1)}]=\pm4\mathcal{G}_{\pm2}^{(1)}\,,\qquad 
[\mathcal{G}_{\mp 2}^{(1)},\mathcal{Q}_a^{0}]=\pm 2(\mathcal{Q}_a^{2}\mp i\epsilon_{ab}\mathcal{Q}_b^{2})\,,&\\
\label{nonlinear4}
&[\mathcal{G}_{-2}^{(1)},\mathcal{G}_{+2}^{(1)}]=8(\mathcal{H}-4)(\mathcal{H}(2\mathcal{H}-9)+\Pi_-
(\mathcal{H}^2-4\mathcal{H}+24))\,,&\\
\label{nonlinear5}
&[\mathcal{G}_{\mp 2}^{(1)},\mathcal{Q}_a^{2}]= \pm 2(-80 + 4 \mathcal{H} + \mathcal{H}^2)(\mathcal{Q}_a^{0}\pm
i\epsilon_{ab}\mathcal{Q}_b^{0})\,,&\\
\label{nonlinear7}
&\{\mathcal{Q}_a^{2},\mathcal{Q}_b^{2}\}=2\delta_{ab}
(\eta+1)(\eta+3)(\eta+7)|_{\eta=\mathcal{H}+2\sigma_3-9}\,,&
\end{eqnarray}
where $\Pi_-=\frac{1}{2}(1-\sigma_3)$.   
The common eigenstates of  $\mathcal{H}$ and $\mathcal{Q}^0_1$ are  
\be
\label{states-3}
\Psi_{n}^{+}=
\left(
\begin{array}{c}
(\mathcal{E}_n)^{-1/2} A_{(-)}^-\psi_{2n+1}   \\
\psi_{2n+1}
\end{array}
\right),
\qquad 
\Psi_{n}^{-}=\sigma_3\Psi_{n}^+,
\ee
where $\mathcal{Q}^0_1\Psi^\pm_n=\pm\sqrt{\mathcal{E}_n}\Psi^\pm_n$,
and we have here the relations 
$\Psi_{n}^\pm=(\mathcal{G}_{+2}^{(1)})^n\Psi_0^{\pm}$ and $\mathcal{G}_{-2}^{(1)}\Psi_0^{\pm}=0$.
As a result one can generate all the complete set of eigenstates of the system 
by applying the generators of superalgebra to any of the two ground states 
$\Psi_0^+$ or $\Psi_0^-$, and therefore the restricted set of
generators we have chosen is the complete spectrum generating 
set for the superextended system (\ref{hamilisodef}).
We also note  here that the action of generators not included in the set 
$\mathcal{U}_{0,2}^{(1)}$ can be  reconstructed by considering the action of $\mathcal{U}_{0,2}^{(1)}$'s elements 
 in the Hilbert space of the system. 
\vskip0.1cm
 
The complete set of (anti)-commutation relations 
 (\ref{nonlinear1})-(\ref{nonlinear7}) corresponds to   a nonlinear deformation of superconformal 
algebra $\mathfrak{osp}(2|2)$. 
The first relation from   (\ref{nonlinear1})
and equation (\ref{nonlinear4})  represent
 a nonlinear  deformation of $\mathfrak{sl}(2,\mathbb{R})$ 
with commutator $[\mathcal{G}_{-2}^{(1)},\mathcal{G}_{+2}^{(1)}]$
 to be a cubic polynomial in $\mathcal{H}$.
{}From the superalgebraic relations it follows that like in the linear case 
of superconformal $\mathfrak{osp}(2|2)$ symmetry 
discussed in  Section \ref{Section3.2},
here the extension of the 
set of generators $\mathcal{H}$, $\mathcal{Q}^0_a$ and $\Sigma$ of 
the $\mathcal{N}=2$ Poincar\'e super-symmetry 
by any one of the dynamical integrals $\mathcal{Q}^2_a$, $a=1,2$,
$\mathcal{G}_{+2}^{(1)}$ or $\mathcal{G}_{-2}^{(1)}$ recovers 
all the complete set of generators of the nonlinearly deformed 
superconformal $\mathfrak{osp}(2|2)$ symmetry.
\vskip0.2cm

Due to a gapless deformation of the AFF model,
here similarly to the case of the non-deformed superconformal $\mathfrak{osp}(2|2)$ symmetry, 
the super-extension
based on the positive scheme 
is characterized by essentially different physical properties.
For the positive scheme  
we take $\lambda_*=3$ in 
(\ref{Hlambda*}),  and identify 
$\mathcal{H}'=\text{diag}\,(L_{(+)}-3,L_{0^+}-3)$
as the extended Hamiltonian. 
This $\mathcal{H}'$  is related to 
$\mathcal{H}$ defined by Eq. (\ref{hamilisodef})
by the equality $\mathcal{H}'=\mathcal{H}-6+4\sigma_3$.
For extended system $\mathcal{H}'$,
supercharges ${\mathcal{Q}'}_a^{0}$ 
have the form  similar to  $\mathcal{Q}_a^{0}$  in (\ref{Q2Cpm2}) 
but  with  $A^\pm_{(-)}$ changed for the  intertwining operators
$A^\pm_{(+)}$. 
Being differential operators of the third order, 
they satisfy relations  
$[\mathcal{H}', {\mathcal{Q}'}_a^{0}]=0$  and 
$\{{\mathcal{Q}'}_a^{0},{\mathcal{Q}'}_b^{0}\}=2\delta_{ab}P_{n_+}(\mathcal{H}'+3)$
with $P_{n_+}(\mathcal{H}'+3)=\mathcal{H}'(\mathcal{H}'-2)(\mathcal{H}'-4)$.
The linear $\mathcal{N}=2$ super-Poincar\'e algebra
of the  system (\ref{hamilisodef}) is changed here for the nonlinearly 
deformed superalgebra with anti-commutator to be  polynomial 
of the third order in Hamiltonian. This system has two nondegenerate states
$(0,\psi_{1})^t$ and $(0,\psi_{3})^t$ of energies, respectively, $0$ and $4$,
and both them are annihilated by both supercharges  ${\mathcal{Q}'}_a^{0}$.
All higher energy levels $\mathcal{E}'_n=4n$ with $n=2,3,\ldots$
are doubly degenerate.
Thus, the nonlinearly deformed $\mathcal{N}=2$ super-Poincar\'e 
symmetry of this system can be identified as partially unbroken \cite{KliPly}
since the supercharges have differential order three but 
annihilate only two nondegenerate physical states. 
Here instead of the spectrum generating set $\mathcal{U}_{0,2}^{(1)}$ formed by   true
and dynamical integrals the same role is played by the set
of integrals 
$\mathcal{U}_{0,2}'^{(1)}=\{\mathcal{H}',\mathcal{G}'^{(1)}_{\pm 2}, \mathbb{I},\sigma_3, {\mathcal{Q}'}_a^{0},
{\mathcal{Q}'}_a^{2}\}$, where fermionic generators are 
${\mathcal{Q}'}_a^{z}=\mathcal{Q}_a^{4-z}$ with $z=0,2$ according with (\ref{frakS}) and (\ref{gencharge}).
Bosonic dynamical integrals  $\mathcal{G}'^{(1)}_{\pm 2}$ are given here by
\begin{eqnarray}
\label{Cpm2Hprime}
\mathcal{G}_{-2}'^{(1)}=
\left(
\begin{array}{cc}
 A_{(+)}^-(a^+)A^+_{(-)}&   0 \\
0 &  (L_{0^+}-1)(a^-)^2   
\end{array}\right),\qquad
\mathcal{G}_{+2}'^{(1)}=(\mathcal{G}_{-2}'^{(1)})^\dagger\,,
\end{eqnarray}
where Eq. (\ref{superC'}) have been used for the case
of the present positive scheme. They are generated via anticommutation of
${\mathcal{Q}'}^0_a$ with ${\mathcal{Q}'}^2_b$.
The set of operators $\mathcal{U}_{0,2}'^{(1)}$ generates the nonlinearly deformed 
superconformal $\mathfrak{osp}(2|2)$ symmetry 
given by superalgebra of the form 
(\ref{nonlinear3})--(\ref{nonlinear7})
but with coefficients to be polynomials of higher order in Hamiltonian
$\mathcal{H}'$
in comparison with the case of the system (\ref{hamilisodef}).

\subsection{Rationally extended harmonic oscillator}\label{SecDefQHO}

The example we discuss in this subsection corresponds to the rational  extension  
of the harmonic oscillator based on the dual schemes $(1,2)\sim (-2)$,
for which  $N=3$. 
Different aspects of this system were extensively studied in  literature 
\cite{CarPly,CarPly2}, but it was not considered  yet in the light of the  
nonlinearly extended super-Schr\"odinger
symmetry we investigate here.

 The Hamiltonian produced via Darboux transformation based on the 
 negative scheme is
\begin{eqnarray}
\label{H_-2}
&L_{(-)}=L_{(-2)}=-\frac{d^2}{dx^2}+x^2+8\frac{2 x^2-1 }{(1 + 2 x^2)^2}-2\,,&
\end{eqnarray}
whose spectrum is
$E_0=-5$, $E_{n+1}=2n+1$, $n=0,1,\ldots$.
In this system a gap of size 6 
separates the ground state energy  from the equidistant part of the spectrum,
 where levels are separated from each other by a distance $\Delta E=2$.
 The pair of ladder operators  of the $\mathfrak{C}$-family  
connects  here the isolated ground state  with the equidistant part of the spectrum,
and together with the ladder operators  $\mathfrak{A}^\pm_1$ they form 
the complete spectrum generating set of operators for the system.
The  intertwining operators of the negative scheme are
\begin{eqnarray}
&A_{(-)}^-\equiv  A_{(-2)}^{-}=\frac{d}{dx}-x-\frac{4x}{2x^2+1},\qquad A_{(-)}^+\equiv 
A_{(-2)}^{+}=(A_{(-2)}^{-})^{\dagger}\,.&
\end{eqnarray}
We also have the  intertwining operators $A_{(+)}^\pm \equiv  A_{(1,2)}^\pm$ 
constructed on the base of the  seed states of the positive scheme $(1,2)$. 
These four operators satisfy  their respective intertwining relations 
of the form (\ref{inter0}), and
 their alternate products (\ref{polyA})  reduce here to polynomials 
 $P_{n_-}(L_{(-)})=L_{(-2)}+5\equiv  H_1$, $P_{n_-}(L)=L+5\equiv H_0$ and
$P_{n_+}(L_{(+)})=(L_{(+)}-3)(L_{(+)}-5)$, $P_{n_+}(L)=(L+3)(L+5)$, where $L=L_0$
is the Hamiltonian operator of the harmonic oscillator, and 
$L_{(+)}$ is the Hamiltonian 
produced by positive  scheme, which is related with  $L_{(-)}$, according to (\ref{L+L-}),
by $L_{(+)}-L_{(-)}=6$. 
Here, the eigenstate $A_{(-2)}^{-}\widetilde{\psi_{-2}}=1/\psi_{-2}$  is the isolated  
ground state   of zero energy of the shifted Hamiltonian operator $H_1$.

The superextended  Hamiltonian and its spectrum are 
\begin{equation}
\label{superdefHO}
\mathcal{H}=
\left(
\begin{array}{cc}
 H_1&    0 \\
0 &  H_0  
\end{array}\right),\qquad \mathcal{E}_{0}=0\,, \qquad \mathcal{E}_{n+1}=2n+6\,,\qquad n=0,1,\ldots\,.
\end{equation}
The ground state of zero energy is non-degenerate and corresponds to the ground state 
$(A_{(-2)}^- \widetilde{\psi_{-2}},0)^t$. Other energy levels 
are doubly degenerate
and correspond to eigenstates of the extended Hamiltonian (\ref{superdefHO}) 
and supercharge $\mathcal{Q}^0_1$, see below\,:
\begin{equation}
\Psi_{n+1}^{+}=
\left(
\begin{array}{c}
  (\mathcal{E}_{n+1})^{-1/2} A_{(-2)}^-\psi_{n}   \\
 \psi_{n}
\end{array}
\right),\qquad
\Psi_{n+1}^{-}=\sigma_3\Psi_{n+1}^{+}\,.
\end{equation} 
Witten's index equals one, and the system (\ref{superdefHO}) 
is characterized by unbroken $\mathcal{N}=2$ Poincar\'e supersymmetry.
Now  we use the construction of Section \ref{gen} to produce generators
of the  extended nonlinearly deformed super-Schr\"odinger symmetry 
of the system. 
Following (\ref{gencharge}) and (\ref{superC}), 
we construct the odd operators 
$\mathcal{Q}_{a}^{z}$ with $z=-2,-1,0,\ldots,5$,  and 
matrix bosonic ladder operators $\mathcal{G}_{\pm k}^{(1)}$ with $k=1,\ldots,5$.  
Also we must consider the operators $\mathcal{G}_{\pm k}^{(0)}$ with $k=1,2$ 
defined in (\ref{G_k}).
To obtain all the
 ingredients, we have to use the version of relation  (\ref{reqgen1}) for this system 
translated to the supersymmetric extension of $\mathfrak{C}_{N+k}^\pm$ which is
\begin{equation}
\label{req}
\mathcal{G}_{\pm(3l+n)}^{(1)}=(-\mathcal{G}_{\pm 3}^{(1)})^l\mathcal{G}_{\pm n}^{(1)} \,,\qquad n=3,4,5\,,\qquad l=0,1,\ldots\,.
\end{equation} 
Then we generate the even part of the superalgebra\,: 
\begin{equation}
\label{slr1}
[\mathcal{H},\mathcal{G}_{\pm n}^{(1)}]=\pm 2n \mathcal{G}_{\pm n}^{(1)}\,,
\qquad
[\mathcal{H},\mathcal{G}_{\pm l}^{(0)}]=\pm 2l\mathcal{G}_{\pm l}^{(0)}\,,
\end{equation}
\begin{equation}
\label{slr3}
[\mathcal{G}_\alpha^{(1)},\mathcal{G}_\beta^{(1)}]=P_{\alpha,\beta} \mathcal{G}_{\alpha+\beta}^{(1)}+
M_{\alpha,\beta}\mathcal{G}_{\alpha+\beta}^{(0)}\,,
\quad 
\alpha,\beta=\pm1,\ldots,\pm5\,,
\end{equation}
\begin{equation}
\label{slr4}
[\mathcal{G}_\alpha^{(0)},\mathcal{G}_\beta^{(1)}]=\Pi_-(F_{\alpha,\beta}\mathcal{G}_{\alpha+\beta}^{(1)}+
N_{\alpha,\beta}\mathcal{G}_{\alpha+\beta}^{(0)})\,,
\quad 
\alpha=1,2\,,\quad\beta=\pm1,\ldots,\pm5\,,
\end{equation}
\begin{equation}
[\mathcal{G}_{-1}^{(0)},\mathcal{G}_{+1}^{(0)}]=2\Pi_-,\qquad [\mathcal{G}_{\pm 1}^{(0)},\mathcal{G}_{\mp 2}^{(0)}]=
\pm6\mathcal{G}_{\pm 1}^{(0)}\,,
\qquad 
[\mathcal{G}_{-2}^{(0)},\mathcal{G}_{+2}^{(0)}]=8\Pi_-(\mathcal{H}-5)\,,
\end{equation}
where we put $\mathcal{G}_0^{(1)}=\mathcal{G}_0^{(0)}=1$ and $P_{\alpha,\beta}$, $F_{\alpha,\beta}$, 
$M_{\alpha,\beta}$ and $N_{\alpha,\beta}$ are some polynomials in $\mathcal{H}$ and $\Pi_-=\frac{1}{2}(1-\sigma_3)$,
some of which are numerical coefficients,  whose explicit form is listed 
in Appendix \ref{list}.
We note that 
in equations (\ref{slr3}) and (\ref{slr4}), the  operators
 $\mathcal{G}_{\pm n}^{(1)}$ with $1<n\leq 7$ can appear, 
where for $n>5$ we use relation (\ref{req}) (admitting  $\mathcal{G}_{\pm3}^{(0)}$ 
as coefficients  in the algebra).  
Additionally we note that  the operators 
$\mathcal{G}_{\pm m}^{(0)}$ with $m>2$ in both equations where they appear
are
 absorbed in generators $\mathcal{G}_{\pm m}^{(1)}$.

For eigenstates we have the relations 
 \begin{eqnarray}
 \label{C3psi}
&\Psi_{3j+k}^\pm= (\mathcal{G}_{+3}^{(1)})^j\Psi_k^\pm\,,
\qquad
\Psi_{0}=\mathcal{G}_{-3}^{(1)}\Psi_1^\pm,\qquad j=1,2,\ldots\,,\qquad k=1,2,3\,,&\\
 \label{C2psi}
&\Psi_{j}^\pm= (\mathcal{G}_{+ 1}^{(1)})^j\Psi_1^\pm\,,
\qquad
\mathcal{G}^{(1)}_{\pm 1}\Psi_{0}=\mathcal{G}_{-1}^{(1)}\Psi_{1}^\pm=0\,.&
\end{eqnarray}
Eq.  (\ref{C3psi}) shows that we can connect the
isolated  ground state with the equidistant part of the 
spectrum using $\mathcal{G}_{\pm 3}^{(1)}$, which  are not spectrum generating operators. 
Eq.  (\ref{C2psi}) indicates  that the states in the  equidistant part of 
the spectrum  can be connected by  $\mathcal{G}_{\pm 1}^{(1)}$, but this part
of the spectrum cannot be connected by them  with  
the ground state. Thus 
we have to use a combination of both pairs of these operators. 
On the other hand, the odd operators $\mathcal{Q}_a^z$ satisfy relations (\ref{SUSY}),
where $\mathbb{P}_0=\mathcal{H}$, and, therefore, 
 we have again the linear $\mathcal{N}=2$ Poincar\'e  supersymmetry
as a subsuperalgebra  generated by $\mathcal{H}$, $\mathcal{Q}^0_a$ and $\Sigma$.
The general anti-commutation structure is given by 
\begin{equation}
\label{susy3}
\{\mathcal{Q}_a^{n},\mathcal{Q}_b^{m}\}=\delta_{ab}(\mathbb{C}_{nm}+(\mathbb{C}_{nm})^{\dagger})+
i\epsilon_{ab}(\mathbb{C}_{nm}-(\mathbb{C}_{nm})^{\dagger})\,,
\end{equation}
where $\mathbb{C}_{n,m}=\mathbb{C}_{n,m}(\mathcal{G}_{|n-m|}^{(1)},\mathcal{G}_{|n-m|}^{(0)})$
 are some linear combinations of the indicated 
 ladder operators with coefficients to be polynomials in 
$\mathcal{H}$, $\mathcal{G}_{\pm3}^{(0)}$ and $\sigma_3$. 
Some of these relations 
define  ladder operators, see Eq. 
(\ref{Cn+kQN}). 
For  $n=N=3$ and $m=-1,-2$  we can use 
(\ref{superC'})  knowing that $\mathcal{Q}_{a}'^{z}=\mathcal{Q}_{a}^{3-z}$,
see subsection \ref{susyextension}. 
For structure of anti-commutation relations with 
other combinations of indexes,  see Appendix \ref{list}.  
To complete the description 
of the generated nonlinear supersymmetric  structure, 
we write down  the commutators between the independent lowering operators and 
supercharges\,:
\begin{eqnarray}
\label{susy4}
&[\mathcal{G}_{-m}^{(1)},\mathcal{Q}_a^{n}]=\mathbb{Q}_{m,n}^{1}(\mathcal{Q}_{a}^{n-m}+i
\epsilon_{ab}\mathcal{Q}_b^{n-m})+\mathbb{Q}_{m,n}^{2}(\mathcal{Q}_{a}^{m+n}-i
\epsilon_{ab}\mathcal{Q}_b^{m+n})\,,&\\
\label{susy5}
&[\mathcal{G}_{-m}^{(0)},\mathcal{Q}_a^{n}]=\mathbb{G}_{m,n}^{1}(\mathcal{Q}_{a}^{n-m}+i
\epsilon_{ab}\mathcal{Q}_b^{n-m})+\mathbb{G}_{m,n}^{2}(\mathcal{Q}_{a}^{m+n}-
i\epsilon_{ab}\mathcal{Q}_b^{m+n})\,.&
\end{eqnarray} 
Here $\mathbb{Q}_{m,n}^{j}$ and $\mathbb{G}_{m,n}^{j}$ with $j=1,2$ are 
polynomials in $\mathcal{H}$  or numerical coefficients, some of which are listed
 in  the sets of general commutation relations in  Appendix \ref{apen-comm}, 
while other are given explicitly in Appendix \ref{list}.
As the odd fermionic operators  are Hermitian, then 
$[\mathcal{G}_{+m}^{(1)},\mathcal{Q}_a^{z} ]=-([\mathcal{G}_{- m}^{(1)},
\mathcal{Q}_a^{z} ])^{\dagger}$, 
and we do not  write them explicitly. 
In matrix language, Eq. (\ref{susy4}) can be written as
\begin{eqnarray}
&[\mathcal{G}_{-m}^{(1)},\mathcal{Q}_a^{n}]= 
\left(
\begin{array}{cc}
  0&  \mathfrak{S}_{n+m}^-  \\
 \mathfrak{S}_{n-m}^+ &  0     
\end{array}
\right)\,,&
\end{eqnarray}    
and an important point here is that  the number $n-m$ could take values less than -2 and $n+m$ could 
be greater than 5,  but fermionic operators are defined with the index $z$ taking integer values
in the interval  $I=[-2,+5]$.  
It is necessary to  remember that we cut the series of $\mathfrak{S}_z^\pm$ 
because operators outside the defined interval are reduced to combinations (products) of other basic operators.
In this way, we formally apply the definition of $\mathfrak{S}_z^\pm$ outside of the indicated
 interval and 
 use the relation in Appendix \ref{apen-red} to show that these ``new" generated operators reduce to combinations 
of operators with index values in the interval $I$  
and of the generators $\mathfrak{C}_{\pm 3}$.  

Finally, the subsets which produce closed subsuperalgebras here are those defined by $\mathcal{U}_{0,z}^{(1)}$
 in (\ref{U1}), with $z=1,\ldots,5$ in addition to $\mathcal{I}_{N,-k}^{(1)}$ given in (\ref{I1}) with $k=1,2$.
 
With respect to the positive scheme, the super-Hamiltonian is given by $\mathcal{H}'=\text{diag}\,
(L_{(+)}-3,L_0-3)$. 
It has two positive energy singlet states  of the form $(0,\psi_n)$ with $n=1,2$;  besides, 
there are  two ground states $\Psi_0^+=(\phi_0,\psi_0)$ and $\Psi_0^-=\sigma_3\Psi_0^+$ of energy $-2$.
According 
to the construction from the previous section, 
the fermionic operators here are $\mathcal{Q}^{'z}_a=\mathcal{Q}^{3-z}_a$, and
the basic subsets which generate 
closed subsuperalgebras are $\mathcal{U}_{0,k}'^{(1)}$  and $\mathcal{I}_{N,l}'^{(1-2\theta(l))}$ 
with $k=3,4,5$ and $l=-1,-2,4,5$.  

One can  note  that considering  $\mathcal{G}_{\pm 3}^{(1)}$ as coefficients, 
the subset $\{\mathcal{H}, \mathcal{G}_{\pm 3}^{(1)},\sigma_3,
 \mathcal{Q}_a^{-2},\mathcal{Q}_a^{1},\mathcal{Q}_a^{4}, \mathbb{I}  \}   $
also generates  a closed nonlinear superalgebraic structure.

\subsection{Further possibilities for super-extensions}\label{SecFigs}

As we have seen in Section \ref{gen}, there is a big variety of Darboux transformations,
performed by the operators (\ref{genA}), which match the (half-) harmonic oscillator 
with a specific rationally deformed system of class $c$, but with different shift constants.
In fact, these operators satisfy the intertwining relations 
$A_{(\pm)}^-(a^\pm)^n L=(L_{(\pm)}\mp 2n)A_{(\pm)}^-(a^\pm)^n$ and
$A_{(\pm)}^-(a^\mp)^n L=(L_{(\pm)}\pm 2n)A_{(\pm)}^-(a^\mp)^n$. 
Such kind of  transformations also allow us to construct an 
$\mathcal{N}=2$ superextended Hamiltonian of the form (\ref{Hlambda*}) by choosing 
$(L_{(\pm)}\pm 2n)$ as $\breve{L}$, and  its spectrum will be 
different from the spectrum of $\mathcal{H}$ and $\mathcal{H}'$. 
We shall not discuss here
the corresponding superconformal structures of such superextended systems, but 
restrict ourselves just by some  general comments related to their spectra. 

\begin{figure}[H]
\begin{minipage}[r][2in][t]{0.5\textwidth}
\begin{center}
\includegraphics[scale=0.22]{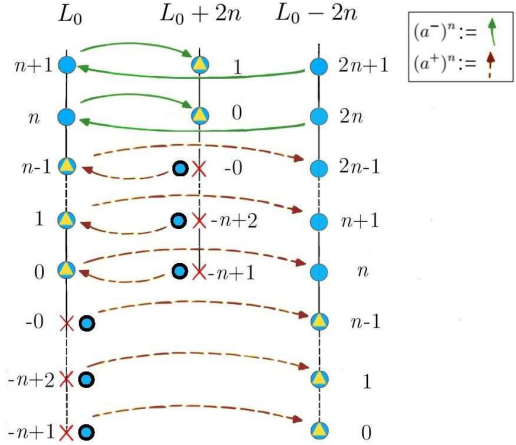} 
\end{center} 
\end{minipage} 
\begin{minipage}[r][2in][t]{3in} 
\caption{\small{The diagram shows the shifted harmonic oscillators $L_0\pm 2n$ 
as super-partners
of $L_0$. 
Symbol  X indicates non-physical seed  states $\psi_{-k}$,
blue discs with boldface border indicate non-physical states 
$\widetilde{\psi_{-k}}$, see Eq. (\ref{secondlinearindependent}),
and circles with  inscribed yellow triangles indicate physical 
seed states $\psi_n$.
 Action of the corresponding intertwining operators is shown by arrows.  } }
 \label{figure1}
\end{minipage}
\end{figure}

\begin{figure}[H]
\begin{center}
\includegraphics[scale=0.6]{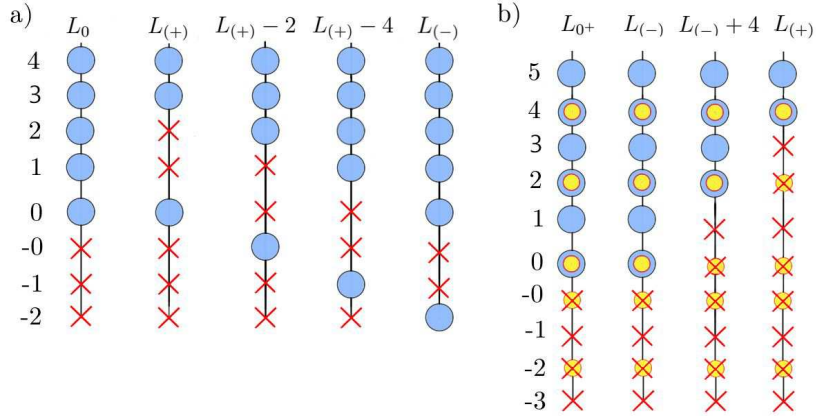} 
\caption{\small{The diagrams show some modifications of the systems discussed in this section.
On the left panel  $a)$, the operator $L_{(+)}=L_{(-)}-6$, and  $L_{(-)}$ is given by (\ref{H_-2}).
On the right panel  $b)$, the operator $L_{(-)}$ is defined in (\ref{L(-3)}), and $L_{(+)}=L_{(-)}-8$. 
Circles with smaller yellow circles inside indicate 
even states which are non-physical
eigenstates;
small yellow crossed circles correspond to
non-physical Wick-rotated even
eigenstates.}} 
\label{figure2}
\end{center} 
\end{figure}

As any DCKA transformation can be decomposed step by step, it is convenient to start 
by considering the mapping performed by intertwining operators 
$(a^\pm)^n$.
Due to the shape 
invariance of the QHO $L_0$, 
the two Darboux schemes $(0,1,\ldots,n-1)$ and 
$(-0,-1,\ldots,-(n-1))$ produce the same but shifted 
Hamiltonian operators $L_0+2n$ and $L_0-2n$, respectively. 
Then, following (\ref{Hlambda*}), we find  that
the superextended systems 
$(L_0+2n,L_0)$ and $(L_0-2n,L_0)$ 
 have similar  spectra. 
This is summarized by the diagram 
on Figure \ref{figure1}, 
\vskip0.7cm
\noindent
{}From the diagram   
it is easy to reed the spectrum of the corresponding superextended system. If 
we chose $\breve{L}=L_0+2n$, the intertwining operator $(a^-)^n$ annihilates the first $n$ physical states
of $L_0$,
and  the superextended system $(L_0,L_0+2n)$
has $n$ singlet eigenstates  given by 
$(\psi_j,0)^t$ with $j=0,\ldots,n-1$. 
On the other hand, taking 
 $(a^+)^n$ as intertwining operator, we generate  the superextended system
$(L_0,L_0-2n)$  characterized by $n$ singlet states   
$(0,\psi_j)^t$. 
In correspondence with these arguments, 
the intertwining operators of the form $A_{(\pm)}^-(a^\pm)^n$ and $A_{(\pm)}^-(a^\mp)^n$ have the following interpretation:
we first shift the initial system  
as much as we want with the corresponding power of $a^\pm$, 
and then we produce the nontrivial deformation. 
As a result, the final superextended system will present 
singlet states in correspondence with this shift. To illustrate the process, consider the two
examples presented diagrammatically  in Figure \ref{figure2}. 
\vskip-0.7cm
\noindent
This diagram follows the same logic of Fig. \ref{figure1},
 but without showing the action of 
intertwining operators. According to  the  left panel a) of Fig. \ref{figure2}, 
one can construct the following 
super-extensions.  By matching  $L_0$ with $L_{(+)}$,
 we have the positive scheme discussed at the end of 
the previous  subsection. 
Matching $L_0$ with $L_{(+)}-2$ results in the system with two singlet states of the form  
$(0,\psi_{n})^t$ with $n=0,1$, and one more singlet of the form  
$(\phi_0,0)^t$, where $\phi_0$ is the ground state of $L_{(+)}$. 
By matching  $L_0$ with $L_{(+)}-4$,
 we obtain the system with singlet  states $(0,\psi_0)$ and
 $(\phi_0,0)^t$.
Finally, by matching $L_{0}$ with $L_{(+)}-6=L_{(-)}$,
 we reproduce the negative scheme. 
In all these cases we use the chain of relations 
$$
\phi_0=A_{(+)}^-(a^+)^3\widetilde{\psi_{-2}}=
A_{(+)}^-(a^+)^2\widetilde{\psi_{-1}}=
A_{(+)}^-(a^+)\widetilde{\psi_{-0}}=A_{(+)}^-\psi_{0}=A_{(-)}^-\widetilde{\psi_{-2}}
$$  
to be equalities modulo numerical multiplicative factors. 
A similar analysis can be done for deformations 
of the $L_1^{iso}$ with the help of the right panel b). 

Any of these shifted deformed systems can be paired 
each other as super-partners.  For example, we can pair
$L_{(+)}$ and $L_{(-)}$ from panel a), 
and the intertwining operator for them 
will be the operator $\mathfrak{C}_{3}^\pm$.  
In this sense, each deformed system of a class $c$ 
can be paired with its proper  copy 
but with the Hamiltonian displaced for additive constant $2cj$,
$j=1,2,\ldots$,
by using  $\mathfrak{A}_{cj}^\pm$,
$\mathfrak{B}_{cj}^\pm$, or $\mathfrak{C}_{cj}^\pm$
as intertwining  operators.

\section{Discussion and outlook}
\label{conclusion}
In conclusion, we indicate  some open problems to be interesting 
for further investigation.

\vskip0.2cm

We  considered super-extensions 
of rationally deformed harmonic oscillator and AFF models 
paired, respectively,  with the undeformed harmonic 
or half-harmonic oscillator systems. 
But one could construct superextended systems composed 
from the pairs of rationally deformed systems having in mind that the corresponding 
dual schemes can be interpreted as periodic Darboux chains \cite{VesSha}, in which
the harmonic oscillator will  just be an intermediate 
system between $L_{(-)}$ and $L_{(+)}$. 
In the corresponding superextended  Hamiltonian 
(\ref{Hlambda*}) we then will have   $\breve{L}=L_{(+)}$ and $L=L_{(-)}$, 
for which the intertwining operators are given by the sets of the ladder operators
of the  three families considered here. 
Such a   generalization was discussed very schematically 
 in Section \ref{SecFigs}.
The use of some singular systems in the middle of the
Darboux chain could allow to construct  intertwining operators of less order. 
Situation like that appears in the case of exotic supersymmetry
in finite-gap systems obtained from the free particle  \cite{AraMatPly}.  
\vskip0.2cm
 
Recently,  the $\mathcal{PT}$-regularized 
systems (\ref{isoham}) with special 
values of the coupling constant $g=n(n+1)$ but without the confining potential term
were studied in \cite{JM1,JM2} together with their rational deformations, which
are intimately related to the Korteweg-de Vries hierarchy of completely 
integrable systems. It was found there that such systems
reveal unusual  properties, in particular, 
in the context of superconformal symmetries.  
It would be interesting to investigate the 
$\mathcal{PT}$-regularized versions 
of the superextended rationally deformed AFF systems studied  by us here.
We note that the 
$\mathcal{PT}$-regularized  version of the undeformed model (\ref{isoham})
was considered earlier in \cite{PT1}. 
The indicated problem also is interesting in another aspect.
If in the $\mathcal{PT}$-regularized AFF model  we reconstruct the frequency 
parameter $\omega$ and consider the limit $\omega\rightarrow0$, 
we  reproduce the completely invisible zero-gap systems with $g=n(n+1)$ investigated in Refs. \cite{JM1,JM2}.
Based on certain ladder operators we studied here, 
one can also reproduce  the higher derivative Lax-Novikov 
integrals which underly peculiar properties  of the systems 
from  \cite{JM1,JM2}. However, in the same na\"ively 
applied limit $\omega\rightarrow 0$ additional terms 
appearing in the rationally deformed versions of the 
AFF model turn into zero. Therefore, the question is 
whether the rational terms responsible, in particular,  
for the origin of the extreme wave solutions in \cite{JM1}
can be reproduced from the rational terms 
we have here by considering  limit
$\omega\rightarrow 0$ in some indirect way.

\vskip0.2cm
 
We were dealing here only with rationally deformed AFF models with special coupling constant
values $g=n(n+1)$.  It would be interesting to extend 
our analysis for the case of rational deformations
of the AFF model of general form (\ref{isoham}).
For this, it will be necessary to generalize   
the three families of the ladder operators  for the case of arbitrary values
of the parameter $\nu$.
Some rational deformations of the  system (\ref{isoham}) with arbitrary coupling constant values 
were considered in \cite{Grandati2}. 
If the technique of dual schemes can be generalized for the case of $\nu\in\R$,
our approach can also be extended for rational deformations
of the AFF model (\ref{isoham}).
Some preliminary considerations indicate that this indeed can be done.
In this context the case of gapless deformations 
is of a special interest  due to their complete isospectrality
to the original system (\ref{isoham}) with corresponding value of the parameter $\nu$, 
which  attracted considerable 
interest in the physics of anyons some time ago \cite{LeiMyr,Poly}.
This also could shed a light on the properties of the 
$\mathcal{PT}$-regularized two-particle Calogero systems 
 with arbitrary values of $g=\nu(\nu+1)$ but without the confining potential term.
One can expect that they have to be essentially different from the peculiar 
properties of the class of such systems 
with $g=n(n+1)$ studied  in \cite{JM1,JM2}.
\vskip0.2cm

Deformed superconformal symmetry we investigated  is
based essentially  on the higher-derivative integrals of motion.
Such integrals are associated usually with the hidden
symmetries  \cite{Cariglia}.  In their study, 
the Eisenhart-Duval lift  \cite{Eisenhart-lift,Duval,DGH,CGGH} 
plays a very important role by allowing to 
look at the system from the perspective of Riemannian geometry 
and to establish relation between different systems
possessing hidden symmetries.
It would be very interesting 
to consider the systems we studied by applying to 
them the method of the Eisenhart-Duval lift.

\vskip0.3cm

\noindent {\large{\bf Acknowledgements} } 
\vskip0.2cm

L.I. acknowledges the CONICYT scholarship 21170053
and the project FCI-PM-02 (USACH).

\appendix 
\section*{Appendix}

\section{Operator identities  (\ref{ide})}
\label{show}

We have
equalities $\ker\, (A_{(-)}^+A_{(+)}^-) = \Delta_+ \cup \tilde{\delta}=\{0,1,\ldots,n\}$,
where $\tilde{\delta}=\{A_{(-)}^+A_{(+)}^-
\widetilde{\psi_{-l_1}},\ldots, \\
A_{(-)}^+A_{(+)}^-\widetilde{\psi_{-l_{n_-}}} \}$,
and relation $\A_{(+)}^+\A_{(-)}^-\varphi_{n}\propto (a^+)^N\varphi_{n}$
following from  (\ref{relation-operators}) is used. The  first identity from (\ref{ide})
follows then from equality $\ker A_{(-)}^+A_{(+)}^-$ =$\ker(a^-)^N$ \cite{CarPly}.

In the second relation in (\ref{ide}), 
functions  $f(\eta)$ and $h(\eta)$ are the polynomials
\begin{eqnarray}
\label{fyh}
&f(\eta)\equiv \prod_{l^-_{k}-n_+<0}(\eta+2l^-_{k}+1),\qquad
h(\eta)\equiv \prod_{\check{n}^-_{k}-n_+\geq 0}(\eta+2\check{n}^-_{k}+1)\,,&
\end{eqnarray}
where $l^-_{k}\in \Delta_-$ and $\check{n}^-_{k}$ are 
the absent states in $\Delta_-$. Using the mirror diagram technique \cite{CIP}, 
we obtain the equality    
$\ker f(L_{(-)})A_{(+)}^-(a^+)^{n_-}=\ker h(L_{(-)})A_{(-)}^-(a^-)^{n_+}$, where 
\begin{eqnarray}
\label{kerA}
\small{\begin{array}{rr}
\ker f(L_{(-)})A_{(+)}^-(a^+)^{n_-}=&\text{span}\{0,\ldots,(n_+-1),-0,\ldots,-(n_--1),\\
                                    &\qquad\qquad\{\widetilde{(\check{n}_i^{+}-n_+)}\},
                                       \{-\widetilde{(\check{n}_j^{-}-n_-)}\}\
                                       \}\,.
\end{array}}
\end{eqnarray} 
Indexes $i$ and $j$ are running here over the absent states of both schemes, provided the conditions 
$\check{n}_{i}^{+}-n_+\geq 0$ and 
$\check{n}_{j}^{-}-n_-\geq 0$ are met. A special case corresponds to the 
positive scheme 
 $\Delta_+=(l_1^+,l_1^++1,\ldots,l_1^++q)$, for which  the dual negative 
scheme is   $\Delta_-=(-(q+1),\ldots,-(q+l_1^+))$. Here  
$n_+=1+q$ and $n_-=l_{1}^+$,
there are no states to construct polynomials (\ref{fyh}),
and  we just put
$f(\eta)=h(\eta)=1$. 
Analogously, there are  eigenstates with tilde in  (\ref{kerA}) in this case. In particular,
if $n_+$ is an even number, then the DCKA transformation will produce 
a deformed harmonic oscillator 
with one gap of size $2(l_1^{-}+q+1)=2N$ in its spectrum, while 
if $q$ is an odd number and 
$l_{1}^+=1$, then we generate a gapless deformation of  $L_{1}^{iso}$ (by introducing the potential 
barrier at $x=0$).

\section{Relations between symmetry generators}\label{apen-red}
We first show explicitly how the three families appear by considering the 
commutators
\begin{eqnarray}\nonumber
\small{\begin{array}{l}
[\mathfrak{C}_{N+l}^-,\mathfrak{A}_{k}^-]=P_{n_-}(\eta)|_{\eta=L_{(+)}}^{\eta=L_{(+)}+2k}\mathfrak{C}_{N+k+l}^-\,,\qquad
[\mathfrak{C}_{N+l}^+,\mathfrak{B}_{k}^+]=P_{n_-}(\eta)|_{\eta=L_{(+)}}^{\eta=L_{(-)}-2l}\mathfrak{C}_{N+k+l}^+\,,\\{} 
[\mathfrak{A}_{k}^+,\mathfrak{C}_{N+l}^-]=(-1)^{n_-}T_k(L_{(-)})\mathfrak{A}_{N+l-k}^--P_{n_-}(L_{(+)})T_{l}(L_{(+)}+2l)\mathfrak{C}_{N+l-k}^-\,,  
\\{} 
 [\mathfrak{B}_{k}^-,\mathfrak{C}_{N+l}^+]=(-1)^{n_-}T_{k}(L_{(+)}+2k)\mathfrak{B}_{N+l-k}^+-
T_{l}(L_{(-)})P_{n_+}(L_{(-)}-2l)\mathfrak{C}_{N+l-k}^+\,,\\ {}
[\mathfrak{C}_{N+k}^+,\mathfrak{C}_{N}^-]=P_{n_+}(L_{(-)}-2k)\mathfrak{A}_{k}^--P_{n_-}(L_{(-)}+2N)\mathfrak{B}_{k}^-\,,
\\{}
[\mathfrak{C}_{N\pm k}^{\pm},\mathfrak{C}_{N\pm l}^{\pm}]=0\,,\qquad\l \geq 0\,,
\end{array}}
\end{eqnarray}
where polynomials $P_{n_\pm}(\eta)$ and $T_k(\eta)$ are defined by Eqs. (\ref{polyA}) and (\ref{Tk}).
These 
commutators should be interpreted as recursive relations which generate the elements of the 
three families of the ladder operators proceeding  from the spectrum generating set 
of operators  with   $l=r(N,c)$ and $k=c$. 
On the other hand, the commutators of the  ladder operators with their own 
conjugate counterparts are
\begin{eqnarray}\small{
\begin{array}{c}
[\mathfrak{A}_k^-,\mathfrak{A}_{k}^+]=P_{n_-}(\eta-2k)P_{n_-}(\eta)T_k(\eta)|_{\eta=L_{(-)}}^{\eta=L_{(-)}+2k}\,,\\\,
[\mathfrak{B}_k^-,\mathfrak{B}_{k}^+]=P_{n_+}(\eta-2k)P_{n_+}(\eta)T_k(\eta)|_{\eta=L_{(+)}}^{\eta=L_{(+)}+2k}\,,\\\,
[\mathfrak{C}_{N\pm k}^-,\mathfrak{C}_{N\pm k}^+]=P_{n_-}(\eta)P_{n_+}(x-2k)T_k(\eta)|_{\eta=L_{(-)}}^{\eta=L_{(+)} \pm 2k}\,.
\end{array}}
\end{eqnarray}
In this way, we obtain a deformation of $sl(2,\R)$ in (\ref{sl2rh}). 

Below we present some relations between lowering ladder operators, from which analogous relations
for  raising operators can be obtained via Hermitian conjugation. 

The definitions of the three families  automatically  provide the following relations: 
\begin{eqnarray}
\label{a-c,b-c}
&\mathfrak{A}_{N+k}^-=(-1)^{n_-}P_{n_-}(L_{(-)})\mathfrak{C}_{N+k}^-\,,\qquad
\mathfrak{B}_{N+k}^-=(-1)^{n_-}\mathfrak{C}_{N+k}^-P_{n_+}(L_{(+)}),&
\\
\label{C-}
&\mathfrak{C}_{N-(N+k)}^-\equiv\mathfrak{C}_{-k}^-=(-1)^{n_-}P_{n_+}(L_{(-)}+2N)\mathfrak{A}_{k}^+\,,&\\
\label{reqgen1} 
&\mathfrak{C}_{2N+l+k}^-=(-1)^{n_-}\mathfrak{C}_{N+l}^-\mathfrak{C}_{N+k}^-\,, & \\
\label{reqgen2} 
&(\mathfrak{C}_{N+k}^-)^2=(-1)^{n_-}\mathfrak{C}_{2N+2k}^-\,,&
\end{eqnarray}
where $k,l=0,1,\ldots$. 
Eq. (\ref{a-c,b-c}) means that operators of families $\mathfrak{A}$ and $\mathfrak{B}$ with index $k\geq N$ 
are essentially the operators of the $\mathfrak{C}$ family.  
Eq. (\ref{C-}) shows that operators of the form $\mathfrak{C}_{-k}^\pm$ are not basic. 
If in  (\ref{reqgen1}) one fixes $l=r(N,c)$, then all the operators with index equal or
greater than  $N+r(N,c)$  reduce to the products  of the basic elements. 
Finally, Eq. (\ref{reqgen2})  
  means that the square of an operator of $\mathfrak{C}$-family with odd index 
 $N+k$ is a physical operator, but not basic. The unique special case is when $c=2$, $N$ is odd, and $k=0$
 since there is no product of physical operators of lower order which
 could make the same job. {}From here we conclude that the basic operators 
 are given by (\ref{genlad}). 
 
For one-gap system we can use the  second equation in (\ref{ide})
(where $f(\eta)=h(\eta)=1$) to find some relations between operators with indexes less than $N$\,:
\begin{eqnarray}
\label{ABTOC}
\small{
\begin{array}{ll}
\mathfrak{C}^-_{n_+-k}=(-1)^{n_-}\mathfrak{A}_{n_+-k}^-T_{k}(L_{(-)})\,,&
\mathfrak{C}^-_{n_--k'}=(-1)^{n_-}\mathfrak{B}_{n_--k'}^-T_{k'}({L_{(-)}}+2(n_+-k'))\,,\\
\mathfrak{A}_{n_++k'}^-=(-1)^{n_-}\mathfrak{C}_{n_++k'}^-T_{k'}(L_{(-)})\,,&
\mathfrak{B}_{n_-+k}^-=(-1)^{n_-}\mathfrak{C}_{n_-+k}^-T_{k}(L_{(-)}+2n_+)\,,
\end{array}}
\end{eqnarray}
where $k=0,\ldots,n_+$ and $k'=0,\ldots,n_-$. 
By considering the ordering relation between $n_-$ and $n_+$, we can combine relations (\ref{ABTOC})
to represent operators of the $\mathfrak{A}$ family in terms of $\mathfrak{B}$ family or vice versa. 
For the case $n_-<n_+$  we have    
\begin{equation}
\label{BcomoA}
\mathfrak{B}_{n_+-k}^-=T_{(n_+-n_--k)}(L_{(-)}+4n_+-2k)T_{k}(L_{(-)}+2n_+)\mathfrak{A}_{n_+-k}^-\,,
\end{equation}
where $k=0,\ldots, n_+-n_-$. In other words, only first  
$n_--1$ operators are basic.
In the case  $n_-=1$, there exist  no basic elements in the  $\mathfrak{B}$-family.
As  examples  corresponding to this observation we  have all the deformations produced 
by a unique non-physical state of the form $\psi_{-n}(x)$.  
On the other hand, in the case $n_+<n_-$ we have
\begin{equation}
\label{AcomoB}
\mathfrak{A}_{n_--k}^-=T_{k}(L_{(-)}+2N)T_{(n_--n_+-k)}(L_{(-)}+2(n_--k))\mathfrak{B}_{n_--k}^-\,,
\end{equation}
where  $k=0,\ldots, n_--n_+$. 
According to this, only first  $n_+-1$ elements cannot be written in terms of the operators of 
 $\mathfrak{B}$ family. 
 The unique case in which there exist no basic elements of the families $\mathfrak{A}$ or $\mathfrak{B}$ 
 is when $n_-=n_+=1$, 
 which corresponds to the shape invariance of the harmonic oscillator. As a final result, the basic elements 
 of the three families are given by (\ref{basicsubsetonegap}).

We consider now the relations  between Darboux generators $A_{(\pm)}^-(a^\pm)^n$
and $A_{(\pm)}^-(a^\mp)^n$.  
Using the first relation in (\ref{ide}) and the definition of operators $\mathfrak{C}_{N+l}^\pm$, we obtain relations 
\begin{eqnarray}
&A_{(-)}^-(a^{-})^{N+l}=(-1)^{n_-}P_{n_-}(L_{(-)})A_{(+)}^-(a^{-})^{l}\,,\quad
A_{(+)}^-(a^{+})^{N+l}= (-1)^{n_-}P_{n_+}(L_{(+)})A_{(-)}^-(a^+)^{l}\,,&\nonumber\\
&A_{(+)}^-(a^-)^{N+l+k}=(-1)^{n_-}\mathfrak{C}_{N+l}^-A_{(+)}^-(a^-)^{k}\,,\quad
A_{(-)}^-(a^+)^{N+l+k}=(-1)^{n_-}\mathfrak{C}_{N+l}^+A_{(-)}^-(a^+)^{k}\,,&\nonumber
\end{eqnarray}
where $k,l=0,1,2,\ldots$. If we fix  $l=r(N,c)$, 
then one finds that the basic elements are just (\ref{genA}).
On the other hand for one gap-systems, with the help of (\ref{ide}) one can obtain relations 
\begin{eqnarray}
\label{Q->S+}
&A_{(-)}^-(a^-)^{n_++k}=(-1)^{n_-}A_{(+)}(a^+)^{n_--k}T_{k}(L)\,,&\\
\label{Q->S+2}
&A_{(+)}^{-}(a^{+})^{n_-+k'}=(-1)^{n_-}A_{(-)}^-(a^-)^{n_+-k'}T_{k'}(L+2k')\,,&
\end{eqnarray}
with $k=0,\ldots, n_-$ and $k'=0,\ldots,n_+$. These relations reduce 
the basic subsets of Darboux generators 
to (\ref{frakS}).

\section{(Anti)-Commutation relations for one-gap system}
\label{apen-comm}

In this Appendix we summarize some (anti)commutation relations  
for one-gap deformations of harmonic oscillator systems. 

For the anticommutator of two fermionic operators
in (\ref{SUSY}) we have
\begin{eqnarray}
\label{hbP}
\mathbb{P}_z=\small{ \left\{ \begin{array}{lcc}
   P_{n_-}(\eta)T_{|z|}(\eta)\big|_{\eta=\mathcal{H}+2|z|\Pi_- + \lambda_{-}} & -N<z\leq0\\\vspace{-0.4cm}
   \\ P_{n_-}(\eta)T_z(\eta+2z)\big|_{\eta=\mathcal{H}-2|z|\Pi_- + \lambda_{-}} & 0<z\leq n_+\\                                                                       \end{array}
   \right. \,,}
 \end{eqnarray} 
 and for the positive scheme  
\begin{eqnarray}
\label{h2bP}
\mathbb{P}'_z= \small{\left\{ \begin{array}{lcc}
  P_{n_+}(\eta)T_{|z|}(\eta-2z)\big|_{\eta=\mathcal{H}'-2|z|\Pi_- + \lambda_{+}} & -N<z\leq 0\\\vspace{-0.4cm}
   \\ P_{n_+}(\eta)T_{z}(\eta)\big|_{\eta=\mathcal{H}'+2|z|\Pi_- + \lambda_{+}}  & 0<z\leq n_-\\                                                                       \end{array}
   \right. \,.}
 \end{eqnarray} 
 By virtue of the relation between dual  schemes, the expression $\mathbb{P}'_{z}(\mathcal{H}',\sigma_3)=\mathbb{P}_{N-z}(\mathcal{H}'+N(1+\sigma_3)-\lambda_-+\lambda_+,\sigma_3)$ helps to complete the set of polynomials.  
 
 For the negative scheme we also have
\begin{equation}
\label{CkQ0}
[\mathcal{G}_{-k}^{(2\theta(z)-1)},\mathcal{Q}_a^0]=\frac{1}{2}P_{n_-}(\eta)|_{\eta=\mathcal{H}+\lambda_{-}}^{\eta=\mathcal{H}+
\lambda_{-}+2|z|}(\mathcal{Q}_a^{z}-(2\theta(z)-1)i\epsilon_{ab}\mathcal{Q}_b^{z})\,,\qquad
\end{equation}
where $ z \in (-N,0)\cup(0,2N)$,
while for the positive scheme, where $\mathcal{Q}_{a}^{'z}=\mathcal{Q}_a^{N-z}$ and $\mathcal{G}_{\pm k}^{'(1)}=\mathcal{G}_{\pm k}^{(1)}$ 
when $k\geq N$,  we have
\begin{equation}
\label{B3}
[\mathcal{G}_{-z}^{'(2\theta(z)-1)},\mathcal{Q}_{a}^{'0}]=\frac{1}{2}P_{n_+}(x)|_{x=\mathcal{H}'+\lambda_{+}}^{x=\mathcal{H}'+
\lambda_{+}+2|z|}(\mathcal{Q}_a^{'z}+(2\theta(z)-1)i\epsilon_{ab}\mathcal{Q}_b^{'z})\,,
\end{equation}
where $ z \in (-N,0)\cup(0,2N)$. On the other hand, for the negative scheme the relation
$[\mathcal{G}_{z}^{(1)},\mathcal{Q}_a^z]=\frac{\mathcal{V}_{z}
(\mathcal{H})}{2}(\mathcal{Q}_a^{0}+i(2\theta(z)-1)\epsilon_{ab}\mathcal{Q}_b^{0})\,$
is valid,
where
\begin{eqnarray}
\label{C-kQk}
\mathcal{V}_{z}= \left\{ \begin{array}{lcc}
   P_{n_-}(\eta)T_z(\eta)\big|^{\eta=\mathcal{H}+\lambda_{-}-2z}_{\eta=\mathcal{H}+\lambda_{-}}\,,& -N<z<0\,,\\\vspace{-0.4cm}
   \\ P_{n_-}(\eta)T_z(\eta+2z)\big|^{\eta=\mathcal{H}+\lambda_{-}}_{\eta=\mathcal{H}+\lambda_{-}-2z} \,,& 0<z\leq n_+\,,\\\vspace{-0.4cm}
\\ P_{n_+}(\eta)T_{N-z}(\eta)\big|^{\eta=\mathcal{H}+\lambda_{-}+2N}_{\eta=\mathcal{H}+\lambda_{-}+2(N-z)}\,,& n_+<z\leq N\,,\\\vspace{-0.4cm}
\\ P_{n_+}(\eta)T_{z}(\eta+2z)\big|^{\eta=\mathcal{H}+\lambda_{-}+2N}_{\eta=\mathcal{H}+\lambda_{-}-2z}\,, & N<z<2N\,.
                                                                       \end{array}
   \right. 
    \end{eqnarray} 
In the positive scheme we have
$[\mathcal{G}_{z}^{'(1)},\mathcal{Q}_a^{'z}]=\frac{\mathcal{V}_{z}'(\mathcal{H}')}{2}(\mathcal{Q}_a^{'0}-i(2\theta(z)-1)\epsilon_{ab}\mathcal{Q}_b^{'0})$, 
where $\mathcal{V}_{z}'(\mathcal{H}')$ are given by
\begin{eqnarray}
\label{C-kQk2}
\mathcal{V}_{z}'= \left\{ \begin{array}{lcc}
   P_{n_+}(\eta)T_{z}(\eta+2z)\big|^{\eta=\mathcal{H}'+\lambda_{+}}_{\eta=\mathcal{H'}+\lambda_{+}-2z}\,,  & -N<z<0\,,\\\vspace{-0.4cm}
   \\P_{n_+}(\eta)T_z(\eta)\big|^{\eta=\mathcal{H}'+\lambda_{+}}_{\eta=\mathcal{H}'+\lambda_{+}}\,, & 0<z\leq n_-\,,\\\vspace{-0.4cm}
\\  P_{n_-}(\eta-2N)T_{N-z}(\eta-2z)\big|^{\eta=\mathcal{H}'+\lambda_{+}+2z}_{\eta=\mathcal{H}'+\lambda_{+}}\,,& n_-<z\leq N\,,\\\vspace{-0.4cm}
\\P_{n_-}(\eta)T_z(\eta)\big|^{\eta=\mathcal{H}'+\lambda_{+}+2z}_{\eta=\mathcal{H}'+\lambda_{+}-2N}\,, & N<z<2N\,.
                                                                       \end{array}
   \right. 
 \end{eqnarray} 
These are the missing relations which prove that the subsets $\mathcal{U}_{0,z}^{(2\theta-1)}$ defined in (\ref{U1}),
 satisfy closed superalgebras independently of choosing  the  scheme.
On the other hand, we can use them
to prove that the subsets  $\mathcal{I}_{N,z}^{(2\theta-1)}$ given in (\ref{I1}) also produce  closed superalgebras. 
Other useful relations are 
\be
\label{Cn+lQN+k}
[\mathcal{G}_{-(N+l)}^{(1)},\mathcal{Q}_a^{N+k}]=\frac{1}{2}T_k(\eta)P_{n_+}(\eta+2k)|_{\eta=\mathcal{H}+ \lambda_{-}}^{\eta=
\mathcal{H}+ \lambda_{-}+2(N-k)}(\mathcal{Q}_a^{l-k}+i\epsilon_{ab}\mathcal{Q}_b^{l-k})\,,
\ee
where $l>k$ and $l-k\leq n_+$. For $l<k$ we have
\be
\label{Cn+lQN+k2}
[\mathcal{G}_{-(N+l)}^{(1)},\mathcal{Q}_a^{N+k}]=P_{n_-}(\mathcal{H}+ 
\lambda_-)(\mathcal{Q}_a^{k-l}+i\epsilon_{ab}\mathcal{Q}_b^{k-l})\,,
\ee
and also we can write $[\mathcal{G}_{\pm(N\pm k)}^{(1)},\mathcal{G}_{\pm (N\pm l)}^{(1)}]=0$ for any values of $k$ and $l$.

\section{List of polynomial functions for Section \ref{SecDefQHO}}
\label{list}
\underline{Eq. (\ref{slr3})}\,: $P_{\alpha,\beta}(\mathcal{H})=-P_{-\alpha,-\beta}(\mathcal{H}-2(\alpha+\beta))$, 
and
\begin{eqnarray}
\nonumber\small{
\begin{array}{lll}
P_{-1,1}=\mathcal{H}(6\mathcal{H}-20)-8\Pi_-(\mathcal{H}-3)\,, &
P_{-1,-2}=-2\mathcal{H}+12(1-\Pi_-)\,, &
P_{-1,+2}=10(\mathcal{H}-4)-12\Pi_-\,,\\
M_{-1,+2}=12\,,&P_{-1,-3}=P_{-2,-3}=-6\,, & P_{-1,+3}=-12\,,\\
M_{-1,+3}=24\,, & P_{-1,-4}=P_{-2,-4}=-8\,, & P_{-1,+4}=16(\mathcal{H}-5-\Pi_-)\,,\\
P_{-1,-5}=P_{-2,-5}=-10\,, & P_{-1,+5}=20(\mathcal{H}-6-\Pi_-)\,,&
\end{array}}
\end{eqnarray}
\vskip-0.51cm
\begin{eqnarray}
\nonumber\small{
\begin{array}{ll}
P_{-2,+2}=(\mathcal{H}-4)[8\mathcal{H}(2\mathcal{H}-7)+\Pi_-(4\mathcal{H}^2-44\mathcal{H}+192)]\,,&
P_{-2,+3}=-18\mathcal{H}+96-4(\mathcal{H}-30)\Pi_-\,,\\
 M_{-2,+3}=M_{-2,+4}=-96\,, &P_{-2,+4}=-2(11\mathcal{H}-\Pi_-)+136\,, \\
P_{-2,+5}=1104-340\mathcal{H}+26\mathcal{H}^2+\Pi_-(576-104\mathcal{H}+10\mathcal{H}^2)\,, &
P_{-3,+4}=24\mathcal{H}-144+\Pi_-(2\mathcal{H}-76)\,,\\ M_{-3,+4}=848\,, &
P_{-3,+5}=30\mathcal{H}-180+\Pi_-(8\mathcal{H}-180)\,,\\ M_{-3,+5}=960\,, \\
P_{-4,+5}=40(32-10\mathcal{H}+\mathcal{H}^2-\Pi_-(7\mathcal{H}-32))\,,& M_{-4,+5}=-5760\,, \\
\end{array}}
\end{eqnarray}
\vskip-0.51cm
\begin{eqnarray}
\nonumber\small{
\begin{array}{l}
P_{-4,+4}=4[7\mathcal{H}^3 - 56\mathcal{H}^2+116\mathcal{H}+32+\Pi_-(\mathcal{H}^3 - 64\mathcal{H}^2+572\mathcal{H}+1472)]\,,\\
P_{-5,+5}=2(320-2848\mathcal{H}+1268\mathcal{H}^2-248\mathcal{H}^3+23\mathcal{H}^4)+
\\\qquad\qquad\qquad 8\Pi_- (10000-6212\mathcal{H}+1492\mathcal{H}^2-187\mathcal{H}^3+12\mathcal{H}^4)\,.
\end{array}}\\
\nonumber\end{eqnarray}
\vskip-0.51cm
\noindent 
\underline{Eq.  (\ref{slr4})}\,: $F_{\alpha,\beta}(\mathcal{H})=-F_{-\alpha,-\beta}(\mathcal{H}-2(\alpha+\beta))$, and 
\vskip-0.51cm
\begin{eqnarray}
\nonumber\small{
\begin{array}{lll}
F_{+1,-1}=F_{-1,+3}=0\,,& N_{+1,-1}=-F_{-1,-2}=-N_{-2,+1}=2\,, &
F_{-2,-1}=F_{-2,-2}=-4\,,\\
 F_{-1,+2}=-N_{-1,+3}=6\,, &F_{-1,+4}= F_{-2,+1}=8\,, & F_{-1,+5}=10\,,\\
N_{-1,+2}=F_{-2,+3}=-12\,, & F_{-2,+4}=-16& N_{-2,+3}=N_{-2,+4}=48\,,\\
F_{-1,+1}=4(\mathcal{H}-3)\,, & F_{-2,+5}=24(\mathcal{H}-7)\,,& F_{-2,+2}=12(\mathcal{H}-4)^2\,,
\end{array}}
\end{eqnarray}
while other elements are zero. 

\noindent 
\underline{Eq.  (\ref{susy3})}\,: 
$\mathbb{C}_{\alpha,\beta}=\mathbb{C}_{\beta,\alpha}$\,, and 
\vskip-0.51cm
\begin{eqnarray}
\nonumber\small{
 \begin{array}{ll}
 \mathbb{C}_{-2,-1}=\mathcal{G}_{+1}^{(1)}(\mathcal{H}-6)+8\mathcal{G}_{+1}^{(0)}\,, &
\mathbb{C}_{-2,0}=\mathcal{G}_{-2}^{(1)}+4\mathcal{G}_{-2}^{(0)}\,,\\
\mathbb{C}_{-2,1}=-(\mathcal{H}-4\Pi_-)\mathcal{G}_{-3}^{(1)}\,,&
\mathbb{C}_{-2,2}=(\mathcal{H}+4\Pi_-)\mathcal{G}_{-4}^{(1)}\,,\\
\mathbb{C}_{-2,5}=\mathcal{G}_{-7}^{(1)}=-\mathcal{G}_{-3}^{(1)}\mathcal{G}_{-4}^{(1)}\,,&
\mathbb{C}_{-1,0}= \mathcal{G}_{-1}^{(1)}+\mathcal{G}_{-1}^{(0)}\,,\\
\mathbb{C}_{-1,1}= -\mathcal{G}_{-2}^{(1)}+2\mathcal{G}_{-2}^{(0)}\,,&
\mathbb{C}_{-1,4}= \mathcal{G}_{-5}^{(1)}\,,\\
 \mathbb{C}_{-1,5}= \mathcal{G}_{-6}^{(1)}=-(\mathcal{G}_{-2}^{(1)})^2\,,&
\mathbb{C}_{1,2}= -(\mathcal{H}-2)(\mathcal{G}_{-1}^{(1)}+6\mathcal{G}_{-1}^{(0)})\,,\\
\mathbb{C}_{1,3}=-\mathcal{G}_{-2}^{(1)}+6\mathcal{G}_{-2}^{(0)}\,,&
\mathbb{C}_{1,4}=(\mathcal{H}-4\sigma_3)\mathcal{G}_{-3}^{(1)}\,,\\
\mathbb{C}_{1,5}=(\mathcal{H}-4-10\Pi_-)\mathcal{G}_{-4}^{(1)}\,,&
\mathbb{C}_{2,3}=(\mathcal{H}-2)\mathcal{G}_{-1}^{(1)}-12(\mathcal{H}-4)\mathcal{G}_{-1}^{(0)}\,,\\
\mathbb{C}_{2,4}=(\mathcal{H}-2)\mathcal{G}_{-2}^{(1)}-16(\mathcal{H}+3)\mathcal{G}_{-2}^{(0)}\,,&
\mathbb{C}_{2,5}=((\mathcal{H}-2)(\mathcal{H}-4)-8(\mathcal{H}-5)\Pi_-)\mathcal{G}_{-3}^{(1)}\,,\\
\mathbb{C}_{3,4}=(\mathcal{H}-2)\mathcal{G}_{-1}^{(1)}-16(\mathcal{H}-5)\mathcal{G}_{-1}^{(0)}\,,&
\mathbb{C}_{3,5}=-(\mathcal{H}+2)\mathcal{G}_{-2}^{(1)}-20(\mathcal{H}-4)\mathcal{G}_{-2}^{(0)}\,,\\
\end{array}}\\\nonumber
\small{\begin{array}{l}
\mathbb{C}_{4,5}=[(\mathcal{H}-2)(\mathcal{H}+6)-2\Pi_-(12\mathcal{H}-117))\mathcal{G}_{-1}^{(1)}-720\mathcal{G}_{-1}^{(0)}\,. 
\end{array}}
\end{eqnarray}

\vskip-0.51cm
\noindent
\underline{Eq.  (\ref{susy4})}\,:
\vskip-0.51cm
\begin{eqnarray}
\nonumber\small{
\begin{array}{llll}
\mathbb{Q}_{1,-2}^{1}=2\,, &  \mathbb{Q}_{1,-2}^{2}=5\,, &
\mathbb{Q}_{2,3}^{1}=7(40-\mathcal{H})\,,&\mathbb{Q}_{2,3}^{2}=-3\,,\\
\mathbb{Q}_{1,-2}^{1}=2\,,& \mathbb{Q}_{1,-2}^{2}=5(\mathcal{H}-4)\,,&\mathbb{Q}_{2,4}^{1}=-10(\mathcal{H}-6)\,,
 & \mathbb{Q}_{2,4}^{2}=-4\,,\\
\mathbb{Q}_{1,-1}^{1}=-1\,, & \mathbb{Q}_{1,-1}^{2}=3\mathcal{H}-10\,,&
\mathbb{Q}_{2,5}^{1}=(336-118\mathcal{H}+11\mathcal{H}^2)\,, & \mathbb{Q}_{2,5}^{2}=5\,,\\
\mathbb{Q}_{1,2}^{1}=5(\mathcal{H}-4)\,, & \mathbb{Q}_{1,2}^{2}=\mathcal{H}\,,&
\mathbb{Q}_{3,1}^{1}=3\,, & \mathbb{Q}_{3,1}^{2}=1\,, \\
\mathbb{Q}_{1,3}^{1}=-10\,, & \mathbb{Q}_{1,3}^{2}=-6\,, & \mathbb{Q}_{3,2}^{1}=-8(\mathcal{H}-3)\,,
& \mathbb{Q}_{3,2}^{2}=4\,,\\
\mathbb{Q}_{1,4}^{1}=7(\mathcal{H}-36)\,, & \mathbb{Q}_{1,4}^{2}=4\,, &
\mathbb{Q}_{4,1}^{1}=2\,, & \mathbb{Q}_{4,1}^{2}=-1\,,\\
\mathbb{Q}_{1,5}^{1}=9(\mathcal{H}-56)\,, & \mathbb{Q}_{1,5}^{2}=5\,,&
\mathbb{Q}_{4,2}^{1}=10(3-\mathcal{H})\,,& \mathbb{Q}_{4,2}^{2}=-5\,,\\
\mathbb{Q}_{2,-2}^{1}=-2\,, & \mathbb{Q}_{2,-2}^{2}=4(\mathcal{H}-4)(2\mathcal{H}-7)\,, &
\mathbb{Q}_{5,1}^{1}=5\,, & \mathbb{Q}_{5,1}^{2}=3\,,\\
\mathbb{Q}_{2,-1}^{1}=-1\,, & \mathbb{Q}_{2,-1}^{2}=5(\mathcal{H}+2)\,,&
\mathbb{Q}_{5,2}^{1}=6(\mathcal{H}-1)\,, & \mathbb{Q}_{5,2}^{2}=3\,.\\
\end{array}}
\end{eqnarray}
\vskip-0.3cm
\noindent
\underline{Eq.  (\ref{susy5})}\,:
\vskip-0.7cm
\begin{eqnarray}
\nonumber\small{
\begin{array}{llll}
\mathbb{G}_{1,-2}^1=-1\,, & \mathbb{G}_{1,-2}^2=(8-\mathcal{H})\,, & \mathbb{G}_{2,-2}^1=-1\,,
 & \mathbb{G}_{2,-2}^2=(\mathcal{H}+8)(\mathcal{H}-6)\,,\\
\mathbb{G}_{1,-1}^1=1\,,  & \mathbb{G}_{1,-1}^2=(4-\mathcal{H})\,, & \mathbb{G}_{2,-1}^1=1\,,
  & \mathbb{G}_{2,-1}^2=4-\mathcal{H}\,,\\
\mathbb{G}_{1,0}^1=\mathbb{G}_{1,0}^2=1\,,& \mathbb{G}_{2,0}^1=-\mathbb{G}_{2,0}^2=1\,, &
\mathbb{G}_{1,1}^1=\mathcal{H}-4\,, & \mathbb{G}_{1,1}^2=-1\,,\\
\mathbb{G}_{2,1}^1=\mathcal{H}-2\,, & \mathbb{G}_{2,1}^2=\mathcal{H}\,, &\mathbb{G}_{1,2}^1=\mathcal{H}-4\,,
&\mathbb{G}_{1,2}^2=\mathcal{H}\,,\\
\mathbb{G}_{2,2}^1=-1\,,&\mathbb{G}_{2,2}^2=\mathcal{H}\,,&\mathbb{G}_{1,3}^1=\mathbb{G}_{1,3}^2=-1\,,
& \mathbb{G}_{2,3}^1=4-\mathcal{H}\,,\\
\mathbb{G}_{2,3}^2=-1\,,&\mathbb{G}_{1,4}^1=\mathcal{H}-4\,,&\mathbb{G}_{1,4}^2=-1\,,&
\mathbb{G}_{2,4}^1=2-\mathcal{H}\,,\\
\mathbb{G}_{2,4}^2=1\,,&\mathbb{G}_{1,5}^1=\mathcal{H}-4\,,
 & \mathbb{G}_{1,5}^2=-1\,,
 &\mathbb{G}_{2,5}^1=(\mathcal{H}-2)(\mathcal{H}-4)\,,\\
\mathbb{G}_{2,5}^2=-1\,.
\end{array}}
\end{eqnarray}

\end{document}